\documentclass[aps,prd,tightenlines,superscriptaddress,nofootinbib,twocolumn]{revtex4-1}

\usepackage{amsmath}
\usepackage{amssymb}
\usepackage[utf8]{inputenc}
\usepackage[T1]{fontenc}
\usepackage{graphicx}
\usepackage{subfigure}
\usepackage{epstopdf}
\usepackage{appendix}
\usepackage[colorlinks=true]{hyperref}
\usepackage{placeins}
\usepackage{longtable}
\usepackage{multirow}
\usepackage{bm,array}

\newcommand{\super}[1]{\ensuremath{^{\textnormal{#1}}}}
\newcommand{\sub}[1]{\ensuremath{_{\textnormal{#1}}}}

\newcommand{\numgrbs}{$129$}
\newcommand{\percentnoredshift}{$\sim$85\%}

\newcommand{\bkprob}{$19.3\%$}

\newcommand{\medexdistlowfreq}{$0.8$\,Mpc}
\newcommand{\medexdisthighfreq}{$0.3$\,Mpc}
\newcommand{\medexdistlowfreqkpc}{$0.8$\,kpc}

\begin{document}

\title{Methods and results of a search for gravitational waves associated with gamma-ray bursts using the GEO\,600, LIGO, and Virgo detectors}

\author{%
J.~Aasi$^{1}$,
B.~P.~Abbott$^{1}$,
R.~Abbott$^{1}$,
T.~Abbott$^{2}$,
M.~R.~Abernathy$^{1}$,
F.~Acernese$^{3,4}$,
K.~Ackley$^{5}$,
C.~Adams$^{6}$,
T.~Adams$^{7}$,
P.~Addesso$^{8}$
% ...
}\noaffiliation

\affiliation {LIGO, California Institute of Technology, Pasadena, CA 91125, USA }
\affiliation {Louisiana State University, Baton Rouge, LA 70803, USA }
\affiliation {Universit\`a di Salerno, Fisciano, I-84084 Salerno, Italy }
\affiliation {INFN, Sezione di Napoli, Complesso Universitario di Monte S.Angelo, I-80126 Napoli, Italy }
\affiliation {University of Florida, Gainesville, FL 32611, USA }
\affiliation {LIGO Livingston Observatory, Livingston, LA 70754, USA }
\affiliation {Cardiff University, Cardiff, CF24 3AA, United Kingdom }
\affiliation {University of Sannio at Benevento, I-82100 Benevento, Italy, and INFN, Sezione di Napoli, I-80100 Napoli, Italy. }
\affiliation {Albert-Einstein-Institut, Max-Planck-Institut f\"ur Gravitationsphysik, D-30167 Hannover, Germany }
\affiliation {Nikhef, Science Park, 1098 XG Amsterdam, The Netherlands }
\affiliation {LIGO, Massachusetts Institute of Technology, Cambridge, MA 02139, USA }
\affiliation {Instituto Nacional de Pesquisas Espaciais, 12227-010 - S\~{a}o Jos\'{e} dos Campos, SP, Brazil }
\affiliation {International Centre for Theoretical Sciences, Tata Institute of Fundamental Research, Bangalore 560012, India. }
\affiliation {Syracuse University, Syracuse, NY 13244, USA }
\affiliation {University of Wisconsin--Milwaukee, Milwaukee, WI 53201, USA }
\affiliation {Leibniz Universit\"at Hannover, D-30167 Hannover, Germany }
\affiliation {Universit\`a di Siena, I-53100 Siena, Italy }
\affiliation {INFN, Sezione di Pisa, I-56127 Pisa, Italy }
\affiliation {Stanford University, Stanford, CA 94305, USA }
\affiliation {The University of Mississippi, University, MS 38677, USA }
\affiliation {California State University Fullerton, Fullerton, CA 92831, USA }
\affiliation {INFN, Sezione di Roma, I-00185 Roma, Italy }
\affiliation {University of Birmingham, Birmingham, B15 2TT, United Kingdom }
\affiliation {Albert-Einstein-Institut, Max-Planck-Institut f\"ur Gravitationsphysik, D-14476 Golm, Germany }
\affiliation {Montana State University, Bozeman, MT 59717, USA }
\affiliation {European Gravitational Observatory (EGO), I-56021 Cascina, Pisa, Italy }
\affiliation {LIGO Hanford Observatory, Richland, WA 99352, USA }
\affiliation {SUPA, University of Glasgow, Glasgow, G12 8QQ, United Kingdom }
\affiliation {APC, AstroParticule et Cosmologie, Universit\'e Paris Diderot, CNRS/IN2P3, CEA/Irfu, Observatoire de Paris, Sorbonne Paris Cit\'e, 10, rue Alice Domon et L\'eonie Duquet, F-75205 Paris Cedex 13, France }
\affiliation {Columbia University, New York, NY 10027, USA }
\affiliation {Universit\`a di Pisa, I-56127 Pisa, Italy }
\affiliation {CAMK-PAN, 00-716 Warsaw, Poland }
\affiliation {Astronomical Observatory Warsaw University, 00-478 Warsaw, Poland }
\affiliation {Universit\`a degli Studi di Genova, I-16146 Genova, Italy }
\affiliation {INFN, Sezione di Genova, I-16146 Genova, Italy }
\affiliation {Faculty of Physics, Lomonosov Moscow State University, Moscow 119991, Russia }
\affiliation {LAL, Universit\'e Paris-Sud, IN2P3/CNRS, F-91898 Orsay, France }
\affiliation {NASA/Goddard Space Flight Center, Greenbelt, MD 20771, USA }
\affiliation {University of Western Australia, Crawley, WA 6009, Australia }
\affiliation {Department of Astrophysics/IMAPP, Radboud University Nijmegen, P.O. Box 9010, 6500 GL Nijmegen, The Netherlands }
\affiliation {Universit\'e Nice-Sophia-Antipolis, CNRS, Observatoire de la C\^ote d'Azur, F-06304 Nice, France }
\affiliation {Institut de Physique de Rennes, CNRS, Universit\'e de Rennes 1, F-35042 Rennes, France }
\affiliation {Laboratoire d'Annecy-le-Vieux de Physique des Particules (LAPP), Universit\'e de Savoie, CNRS/IN2P3, F-74941 Annecy-le-Vieux, France }
\affiliation {Washington State University, Pullman, WA 99164, USA }
\affiliation {Inter-University Centre for Astronomy and Astrophysics, Pune - 411007, India }
\affiliation {INFN, Sezione di Perugia, I-06123 Perugia, Italy }
\affiliation {Yukawa Institute for Theoretical Physics, Kyoto University, Kyoto 606-8502, Japan }
\affiliation {Universit\`a degli Studi di Urbino 'Carlo Bo', I-61029 Urbino, Italy }
\affiliation {INFN, Sezione di Firenze, I-50019 Sesto Fiorentino, Firenze, Italy }
\affiliation {University of Oregon, Eugene, OR 97403, USA }
\affiliation {Laboratoire Kastler Brossel, ENS, CNRS, UPMC, Universit\'e Pierre et Marie Curie, F-75005 Paris, France }
\affiliation {VU University Amsterdam, 1081 HV Amsterdam, The Netherlands }
\affiliation {University of Maryland, College Park, MD 20742, USA }
\affiliation {University of Massachusetts Amherst, Amherst, MA 01003, USA }
\affiliation {Laboratoire des Mat\'eriaux Avanc\'es (LMA), IN2P3/CNRS, Universit\'e de Lyon, F-69622 Villeurbanne, Lyon, France }
\affiliation {Universitat de les Illes Balears, E-07122 Palma de Mallorca, Spain }
\affiliation {Universit\`a di Napoli 'Federico II', Complesso Universitario di Monte S.Angelo, I-80126 Napoli, Italy }
\affiliation {Canadian Institute for Theoretical Astrophysics, University of Toronto, Toronto, Ontario, M5S 3H8, Canada }
\affiliation {Tsinghua University, Beijing 100084, China }
\affiliation {University of Michigan, Ann Arbor, MI 48109, USA }
\affiliation {INFN, Sezione di Roma Tor Vergata, I-00133 Roma, Italy }
\affiliation {National Tsing Hua University, Hsinchu Taiwan 300 }
\affiliation {Charles Sturt University, Wagga Wagga, NSW 2678, Australia }
\affiliation {Caltech-CaRT, Pasadena, CA 91125, USA }
\affiliation {Pusan National University, Busan 609-735, Korea }
\affiliation {Australian National University, Canberra, ACT 0200, Australia }
\affiliation {Carleton College, Northfield, MN 55057, USA }
\affiliation {Universit\`a di Roma Tor Vergata, I-00133 Roma, Italy }
\affiliation {INFN, Gran Sasso Science Institute, I-67100 L'Aquila, Italy }
\affiliation {Universit\`a di Roma 'La Sapienza', I-00185 Roma, Italy }
\affiliation {University of Brussels, Brussels 1050 Belgium }
\affiliation {Sonoma State University, Rohnert Park, CA 94928, USA }
\affiliation {The George Washington University, Washington, DC 20052, USA }
\affiliation {University of Cambridge, Cambridge, CB2 1TN, United Kingdom }
\affiliation {University of Minnesota, Minneapolis, MN 55455, USA }
\affiliation {The University of Texas at Brownsville, Brownsville, TX 78520, USA }
\affiliation {The University of Sheffield, Sheffield S10 2TN, United Kingdom }
\affiliation {Wigner RCP, RMKI, H-1121 Budapest, Konkoly Thege Mikl\'os \'ut 29-33, Hungary }
\affiliation {Argentinian Gravitational Wave Group, Cordoba Cordoba 5000, Argentina }
\affiliation {Universit\`a di Trento, I-38050 Povo, Trento, Italy }
\affiliation {INFN, Gruppo Collegato di Trento, I-38050 Povo, Trento, Italy }
\affiliation {Northwestern University, Evanston, IL 60208, USA }
\affiliation {Montclair State University, Montclair, NJ 07043, USA }
\affiliation {The Pennsylvania State University, University Park, PA 16802, USA }
\affiliation {MTA E\"otv\"os University, `Lendulet' A. R. G., Budapest 1117, Hungary }
\affiliation {Universit\`a di Perugia, I-06123 Perugia, Italy }
\affiliation {Rutherford Appleton Laboratory, HSIC, Chilton, Didcot, Oxon, OX11 0QX, United Kingdom }
\affiliation {Embry-Riddle Aeronautical University, Prescott, AZ 86301, USA }
\affiliation {Seoul National University, Seoul 151-742, Korea }
\affiliation {Perimeter Institute for Theoretical Physics, Waterloo, Ontario, N2L 2Y5, Canada }
\affiliation {American University, Washington, DC 20016, USA }
\affiliation {College of William and Mary, Williamsburg, VA 23187, USA }
\affiliation {University of Adelaide, Adelaide, SA 5005, Australia }
\affiliation {Raman Research Institute, Bangalore, Karnataka 560080, India }
\affiliation {Korea Institute of Science and Technology Information, Daejeon 305-806, Korea }
\affiliation {Bia{\l }ystok University, 15-424 Bia{\l }ystok, Poland }
\affiliation {University of Southampton, Southampton, SO17 1BJ, United Kingdom }
\affiliation {IISER-TVM, CET Campus, Trivandrum Kerala 695016, India }
\affiliation {Institute of Applied Physics, Nizhny Novgorod, 603950, Russia }
\affiliation {Hanyang University, Seoul 133-791, Korea }
\affiliation {NCBJ, 05-400 \'Swierk-Otwock, Poland }
\affiliation {IM-PAN, 00-956 Warsaw, Poland }
\affiliation {Institute for Plasma Research, Bhat, Gandhinagar 382428, India }
\affiliation {The University of Melbourne, Parkville, VIC 3010, Australia }
\affiliation {INFN, Sezione di Padova, I-35131 Padova, Italy }
\affiliation {Monash University, Victoria 3800, Australia }
\affiliation {SUPA, University of Strathclyde, Glasgow, G1 1XQ, United Kingdom }
\affiliation {Louisiana Tech University, Ruston, LA 71272, USA }
\affiliation {ESPCI, CNRS, F-75005 Paris, France }
\affiliation {Universit\`a di Camerino, Dipartimento di Fisica, I-62032 Camerino, Italy }
\affiliation {Southern University and A\&M College, Baton Rouge, LA 70813, USA }
\affiliation {IISER-Kolkata, Mohanpur, West Bengal 741252, India }
\affiliation {National Institute for Mathematical Sciences, Daejeon 305-390, Korea }
\affiliation {Hobart and William Smith Colleges, Geneva, NY 14456, USA }
\affiliation {RRCAT, Indore MP 452013, India }
\affiliation {Tata Institute for Fundamental Research, Mumbai 400005, India }
\affiliation {SUPA, University of the West of Scotland, Paisley, PA1 2BE, United Kingdom }
\affiliation {Institute of Astronomy, 65-265 Zielona G\'ora, Poland }
\affiliation {Indian Institute of Technology, Gandhinagar Ahmedabad Gujarat 382424, India }
\affiliation {Instituto de F\'\i sica Te\'orica, Univ. Estadual Paulista/ICTP South American Institute for Fundamental Research, S\~ao Paulo SP 01140-070, Brazil }
\affiliation {Andrews University, Berrien Springs, MI 49104, USA }
\affiliation {Trinity University, San Antonio, TX 78212, USA }
\affiliation {University of Washington, Seattle, WA 98195, USA }
\affiliation {Rochester Institute of Technology, Rochester, NY 14623, USA }
\affiliation {Southeastern Louisiana University, Hammond, LA 70402, USA }
\affiliation {Abilene Christian University, Abilene, TX 79699, USA }

\begin{abstract}
In this paper we report on a search for short-duration gravitational wave bursts in the frequency range $64$\,Hz--$1792$\,Hz associated with gamma-ray bursts (GRBs), using data from GEO\,600 and one of the LIGO or Virgo detectors.
We introduce the method of a linear search grid to analyse GRB events with large sky localisation uncertainties, for example the localisations provided by the \emph{Fermi} Gamma-ray Burst Monitor (GBM).
Coherent searches for gravitational waves (GWs) can be computationally intensive when the GRB sky position is not well-localised, due to the corrections required for the difference in arrival time between detectors.
Using a linear search grid we are able to reduce the computational cost of the analysis by a factor of $\mathcal{O}$(10) for GBM events.
Furthermore, we demonstrate that our analysis pipeline can improve upon the sky localisation of GRBs detected by the GBM, if a high-frequency GW signal is observed in coincidence.
We use the method of the linear grid in a search for GWs associated with \numgrbs{} GRBs observed satellite-based gamma-ray experiments between 2006 and 2011.
The GRBs in our sample had not been previously analysed for GW counterparts.
A fraction of our GRB events are analysed using data from GEO\,600 while the detector was using squeezed-light states to improve its sensitivity; this is the first search for GWs using data from a squeezed-light interferometric observatory.
We find no evidence for GW signals, either with any individual GRB in this sample or with the population as a whole.
For each GRB we place lower bounds on the distance to the progenitor, under an assumption of a fixed GW emission energy of $10^{-2}\,\mathrm{M_{\odot}c^{2}}$, with a median exclusion distance of \medexdistlowfreq{} for emission at 500\,Hz and \medexdisthighfreq{} at 1\,kHz.
The reduced computational cost associated with a linear search grid will enable rapid searches for GWs associated with \emph{Fermi} GBM events once the Advanced LIGO and Virgo detectors begin operation.
\end{abstract}

\maketitle

\section{Introduction}
\label{sec:intro}

Gamma-ray bursts (GRBs) are intense flashes of high-energy photons which are observed approximately once per day and are distributed isotropically on the sky \cite{Meszaros:2006gr}.
Since their public discovery in 1973 \cite{Klebesadel:1973co}, it has been found that most GRBs occur at extra-galactic distances, and there is growing evidence that they emit gamma-rays in tightly beamed relativistic jets \cite{Frail:2001wl, FongJetBreak:2012}.
GRBs are grouped into two broad classes based on their spectral hardness and the duration of the initial gamma-ray flash \cite{Kouveliotou:1993cl}.
The progenitors of long-soft GRBs are generally accepted to be core-collapse supernovae (CCSN) in massive, rapidly rotating stars \cite{Galama:1998eb, Bloom:1999af, Woosley:2006sg}.
The progenitors of short-hard GRBs have yet to be definitively constrained by observation, but are widely thought to be associated with the mergers of binary neutron star or neutron star-black hole systems.
Events of this sort are referred to as compact binary coalescences (CBCs) \cite{Eichler:1989nu, Narayan:1992dt, Mochkovitch:1993cj, Nakar:2007sh, Gehrels:2009sw, Rezzolla:2011ml, Tanvir:2013kn, Berger:2013kn}.
Both the CCSN and CBC scenarios result in the formation of a stellar-mass black hole or magnetar with an accretion disk.
CBCs are expected to be bright sources of gravitational waves (GWs), while the GW emission by CCSN is more speculative.

Although it is expected that most GRB progenitors will be at distances too large for any counterpart GW signals to be detectable by the current generation of observatories, it is possible that a few GRBs could be located nearby.
The non-detection of GW counterparts to GRB\,051103 \cite{Abadie:2012vw} and GRB\,070201 \cite{Abbott:2008kg}, which were short-duration GRBs with error boxes overlapping the M81 galaxy (at 3.6\,Mpc) and the Andromeda galaxy (at 770\,kpc) respectively, ruled out CBC progenitors in M81 or M31 with high confidence.
Studies of long GRBs indicate the existence of a local population of under-luminous events with an observed rate density approximately $10^3$ times that of the high-luminosity population \cite{Soderberg:2006xr, Liang:2007ll, Le:2007sw, Chapman:2007lu, Virgili:2009ll, Howell:2013re}.
Approximately \percentnoredshift{} of the GRBs in our sample do not have measured redshifts, so it is possible that one or more events could be much closer than the typical $\sim\,$Gpc distances of GRBs.

Previously, searches for GWs associated with GRBs were performed on approximately $500$ GRBs which occurred during times when at least two of the LIGO and Virgo detectors were collecting data \cite{PhysRevD.77.062004, Abadie:2010sg, Abbott:2008kg, Abadie:2012br, Abadie:2012vw}.
Related analyses have searched for extended-duration GW signals associated with long GRBs \cite{StampGRB:2013}, and for GWs arising from the oscillation of neutron star $f$-modes in magnetars and soft gamma-ray repeaters \cite{S5Hyperflare:2007, S5SGR:2009, S5A5Magnetar:2011}.
Most recently, data from the fifth and sixth LIGO science runs (S5, S6) and the first, second and third Virgo science runs (VSR1, 2 and 3) were searched for short-duration GW bursts and for signals from CBCs, using GRB events detected by the InterPlanetary Network (IPN) \cite{S5S6IPNGRB:2014}.
No evidence for a signal was found in these searches.

The GEO\,600 detector has its best sensitivity at frequencies greater than $500$\,Hz, and because of its limited sensitivity at low frequencies it was not included in previous searches.
However, the high duty cycle of GEO\,600 yields a substantial number of GRBs which occurred during times of joint observation with one of the LIGO/Virgo detectors.
It is these GRBs which we consider here.

The central engines of GRBs are expected to emit GWs at frequencies above $500$\,Hz through a variety of mechanisms, although the amplitude of this emission is not well constrained by simulation.
Coherent searches for high-frequency signals from GRBs can be computationally challenging when the uncertainty in the GRB sky location is very large, due to the shift in the arrival time of GW signals at widely-separated detectors as the sky position varies across the error box.
Previous searches for GWs associated with poorly-localised GRBs have used a search band of 64-500\,Hz, which preserves the detection efficiency of the search for the frequencies with the best detector sensitivity.

In this paper, we present the methods and results of a search for generic GW burst signals associated with \numgrbs{} GRBs which were detected by satellite-based gamma-ray experiments between February 4\super{th} 2006 and November $3\super{rd}$ 2011, and occurred when GEO\,600 and one other km-scale GW observatory were taking data.
The search targets GW signals with durations $\lesssim1$\,s and frequencies between $64$\,Hz and $1792$\,Hz.
Unlike previous GW searches that analysed GRBs with large sky location uncertainty, we do not repeat the search across the entire GRB uncertainty region.
Instead, we use a linear grid of search points to cover the uncertainty region in the direction of the maximum gradient of the time delay between the detectors.
For a search using data from two widely-separated GW observatories, this technique is sufficient to preserve the sensitivity of the analysis while reducing the computational cost of analysing each GRB by a factor of ten or more.

This paper is organised in the following way.
In Sec.~\ref{sec:gw_signals}, we discuss possible GW signals associated with GRBs that are detectable given our analysis parameters.
In Sec.~\ref{sec:gw_detectors} we describe the GW observatories whose data are used in the search.
Sec.~\ref{sec:methods} describes the methods of the search and the analysis procedure for GW signals, and in Sec.~\ref{sec:results} we describe the sample of GRBs included in our search and present the results.
Finally, in Sec.~\ref{sec:summary} we summarize the analysis and discuss the application of these methods to searches in the era of advanced GW detectors.

\section{Models for GW Signals from GRB Progenitors}
\label{sec:gw_signals}

The progenitors of GRBs are expected to have characteristic GW signatures \cite{Corsi:2012xn} depending on the physics of the GRB central engine.
In this section we describe plausible models for GW emission from these systems, especially at high frequencies ($>500\,\mathrm{Hz}$) where this search is most sensitive.

Short GRBs are believed to be associated with the merger of a neutron star either with another neutron star or a black hole \cite{Berger:2014}.
The inspiral phase of these mergers is expected to be a bright source of gravitational radiation \cite{Thorne:1987gr}.
While most of the GW energy flux from the inspiral occurs at frequencies below $500\,\mathrm{Hz}$, numerical simulations of binary neutron star mergers have shown that substantial GW emission can occur at frequencies greater than $1\,\mathrm{kHz}$ \cite{Kiuchi:2009lt, Sekiguchi2011bn}.
Binary neutron star mergers may result in the formation of a hyper-massive neutron star, which can produce strong GW emission as it collapses to a black hole \cite{Oechslin:2007mp, Hotokezaka:2013re}.

Short GRBs with unknown red shifts could also be produced by giant flares from a local population of soft gamma-ray repeaters \cite{Duncan:1992ns, Tanvir:2005lu, Hurley:2010sh, Mazets:2008gf}.
Sources of this kind are expected to produce some GW energy ($\lesssim 10^{-8}\,\mathrm{M_{\odot}c^{2}}$) in the $1\,\mathrm{kHz}$--$2\,\mathrm{kHz}$ range \cite{CorsiMagnetar:2011, ZinkGMF:2012, Lasky:2012, Ciolfi:2013}.

The progenitors of long GRBs are CCSN in rapidly rotating massive stars.
Simulations of CCSN indicate several methods for GW emission at frequencies of several hundred Hz to $1\,\mathrm{kHz}$, but the amplitude of the emission is highly uncertain \cite{Ott:2009sn}.
The most optimistic emission models arise from the pulsations of a proto-neutron star core, which may release $10^{-7}\,\mathrm{M_{\odot}c^{2}}$ in GWs in a narrow frequency band around $1\,\mathrm{kHz}$ \cite{Ott:2006gw, Ott:2009gw}.

Both types of GRBs are expected to result in a compact object (a neutron star or black hole) with an accretion disk.
Instabilities in the accretion disk can emit significant energy in GWs.
Various semi-analytical scenarios have been proposed which release up to $10^{-2}\,\mathrm{M_{\odot}c^{2}}$--$10^{-1}\,\mathrm{M_{\odot}}c^{2}$ in GWs, all of which correspond to the development of rotational instabilities in the accretion disk or central engine.
Bar mode instabilities, in the $l=2$, $m=2$ non-axisymmetric mode, are optimistic models for GW emission in CCSN; typical frequencies are between $500\,\mathrm{Hz}$ and $2\,\mathrm{kHz}$ \cite{Fryer:2002cc, Kobayashi:2003gr}.
If the deformation remains coherent for $\sim 100\,\mathrm{ms}$, $E\sub{GW} \sim 0.1\,\mathrm{M_{\odot}c^{2}}$ could be emitted at $1\,\mathrm{kHz}$.
An accretion disk that cools rapidly enough to become self-gravitating may fragment into one or more smaller bodies and generate an inspiral-like signal that persists to higher frequencies \cite{Piro:2007co}.
The in fall of matter from a rapidly-rotating accretion disk could generate non-axisymmetric instabilities in a neutron star and produce GWs in the $\sim 700\,\mathrm{Hz}$--$2.4\,\mathrm{kHz}$ band for many seconds following the merger \cite{Piro:2012fa}.
Instead of an accretion disk, a torus may form around the black hole and convert the spin energy of the black hole into GWs in the $1\,\mathrm{kHz}$--$2\,\mathrm{kHz}$ band \cite{vanPutten:2001to,vanPutten:2004ob}.
Numerical simulations have produced similar signals \cite{Shibata:2008me, Kiuchi:2009lt}.

Finally, the oscillation of quasi-normal modes of a hyper-massive neutron star or perturbed black hole can emit GWs with large amplitudes \cite{KokkotasQNM:1999, ShibataRingdown:2006, BausweinRingdown:2012}, although the peak emission is typically outside the search band used in this analysis.

\section{GW Observatories}
\label{sec:gw_detectors}

The GEO\,600 detector (G1), located near Hannover, Germany, is a dual-recycled Michelson interferometer with single-folded arms $600$\,m in length \cite{Grote:2008zz,Grote:2010tg}.
GEO\,600 implements a number of advanced interferometric techniques such as signal recycling and squeezed light to improve sensitivity at frequencies above a few hundred Hz \cite{Abadie:2011dj, GEOSqueeze:2013}.
The LIGO \cite{Abbott:2009li} and Virgo \cite{Accadia:2012vi} observatories are power-recycled interferometers of similar design, with Fabry-Perot cavities in the arms to increase the effective arm length and improve the sensitivity to GWs.
There are two LIGO observatories, located in Hanford, WA, USA and Livingston, LA, USA \cite{Abbott:2009li}.
The Hanford site housed two interferometers, one with $4$\,km long arms, which is referred to as H1, and another with $2$\,km long arms which is referred to as H2.
The H2 instrument ceased data-taking operations in July 2009.
The Livingston observatory has a single interferometer with $4$\,km long arms, referred to as L1.
The Virgo detector, known as V1, has $3$\,km long arms and is located in Cascina, Italy \cite{Accadia:2012vi}.

The GEO\,600 detector has been operated with high duty cycle since 2006, with occasional short breaks for invasive configuration changes and instrumental upgrades.
The LIGO and Virgo observatories have taken data in a series of science runs, during which the detector is kept in its most sensitive state, separated by periods of intense commissioning activity.

The fifth LIGO science run (S5) started on November 1\super{st} 2005 and ended on October 1\super{st} 2007.
During S5, the H1, H2, and L1 interferometers operated near their design sensitivity, with duty cycles of approximately 70\%.
The H2 interferometer continued to collect science data on an opportunistic basis from October 1\super{st} 2007 to June 1\super{st} 2009, during a period of instrumental upgrades to the H1 and L1 detectors.
The sixth LIGO science run (S6) was held from July 7\super{th} 2009 to October 20\super{th} 2010.
In S6, the H1 and L1 were operated with duty cycles of $52\%$ and $47\%$ respectively, and both surpassed their sensitivities from S5.

The first Virgo science run (VSR1) started on May 18\super{th} 2007 and ended on October 1\super{st} 2007.
The second Virgo science run (VSR2) was held from July 7\super{th} 2009 to January 8\super{th} 2010, and the third Virgo science run (VSR3) was held from August 11\super{th} 2010 to October 19\super{th} 2010.
The fourth Virgo science run (VSR4) was held from May 20\super{th} 2011 to September 5\super{th} 2011, followed by a period of opportunistic data collection that ended on November 3\super{rd} 2011.
Virgo's duty cycle was 71\% for VSR2-4.

\begin{figure}
  \centering
  \includegraphics[width=0.5\textwidth]{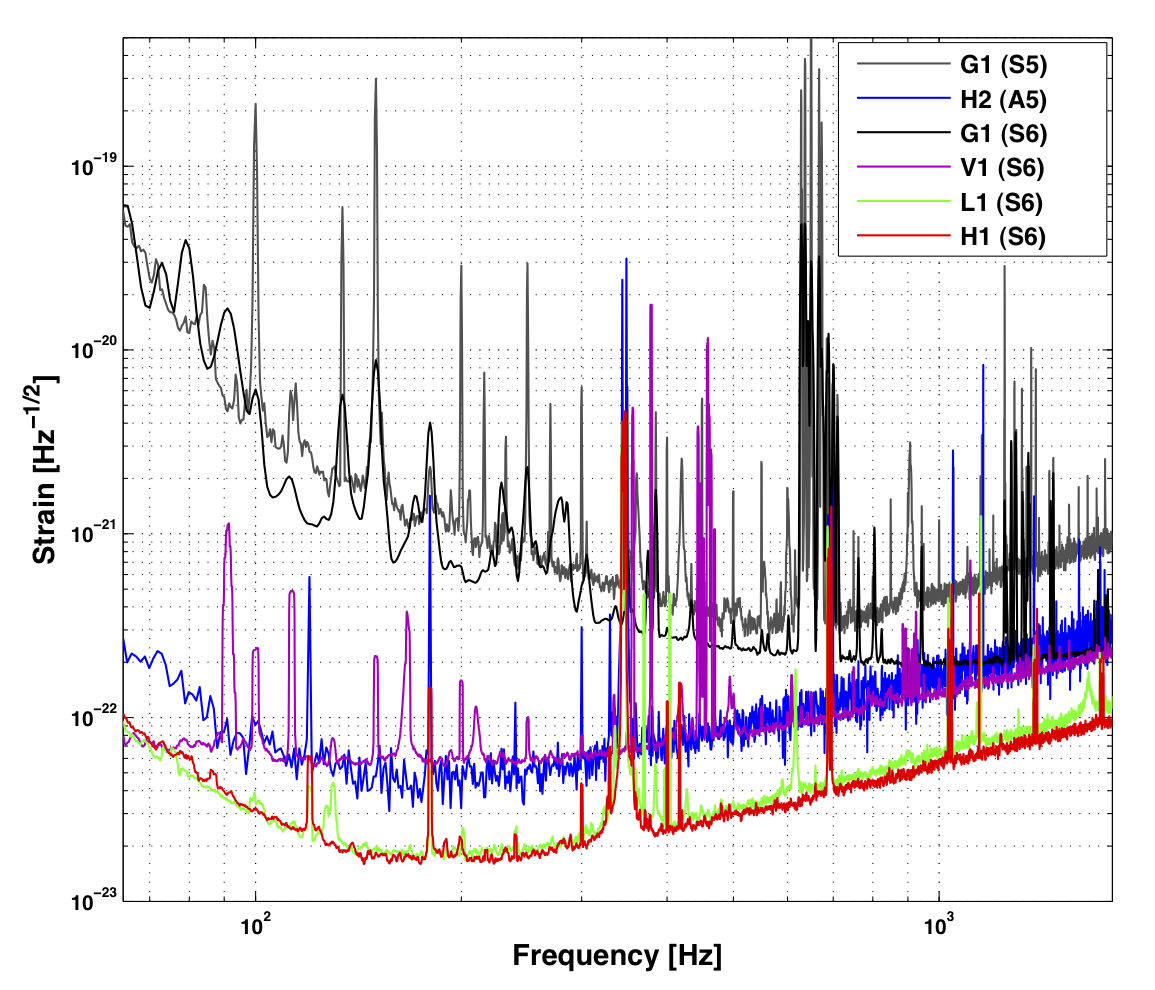}
  \caption{Strain sensitivity for the detectors used in this analysis.
           GEO\,600 is shown at two epochs from the S5 and S6 science runs.
           (A colour version of this figure is available online.)}
  \label{fig:IFO_sensitivity}
\end{figure}

Fig.~\ref{fig:IFO_sensitivity} shows representative sensitivity curves, in terms of amplitude spectral density, of the GEO\,600, LIGO, and Virgo interferometers during these science runs.

\section{Search Methods}
\label{sec:methods}

The coherent analysis algorithm used in this search is \textsc{X-Pipeline} \cite{Sutton:2009gi}, and the overall search procedure follows that used in previous searches for generic GW signals during the S5-VSR1 \cite{Abadie:2010sg} and S6-VSR2,3 \cite{Abadie:2012br} science runs\,\footnote{The most recent searches for short-duration GWs associated with GRBs have included a modelled search for CBC signals (either NS-NS or NS-BH) on the short GRBs included in the sample \cite{Abadie:2012br, S5S6IPNGRB:2014}.
Due to the sensitivity of the GEO\,600 detector at low frequencies (where most of the power of a CBC waveform is emitted) we did not perform a dedicated search for CBCs as part of our analysis.}.
Following the method of the S6-VSR2,3 search, we employ a circularly polarised GW signal model; this is motivated by our expectation that the rotation axis of the GRB central engine is likely pointed at the observer, to within $\sim 10^{\circ}$.
In this section we give a brief description of the analysis pipeline and introduce new techniques to accommodate the high frequency sensitivity of the GEO\,600 detector and the sky localisation uncertainties of the \emph{Fermi} GBM \cite{Meegan:2009fe}.

\subsection{Analysis Procedure}

Data from GW detectors surrounding the time of a GRB are divided into an \emph{off-source window}, which is used to characterise the background of transient signals around the time of the GRB, and an \emph{on-source window}, which is searched for GW signals.
To allow for possible GW precursors from, for example, the CCSN associated with long GRBs, the on-source window is [-600,+60]\,s around the onset time of the GRB event.
For very long-lasting GRBs, the on-source window is extended to include the entire T$_{90}$ time of the event, defined as the time interval over which $90\%$ of the total background-subtracted photon counts are observed.
The standard off-source window is $\pm 1.5$\,hours around the time of the GRB trigger, excluding the on-source window.

In the analysis, the time-series data from each detector are whitened using linear predictor error filters; this removes periodic signals (for example from mechanical resonances in the detector) which are present over long time scales.
The detector data streams are time-shifted to synchronise the arrival time of a GW signal incident from the sky position of the GRB.
Discrete Fourier transforms are performed for various time resolutions, ranging from $1/4$\,s to $1/256$\,s, to maximise the search sensitivity to GW bursts of different duration.
For each detector data stream, and each resolution, the time-frequency maps generated by the DFT are weighted by the sensitivity of the detector to the \emph{plus} and \emph{cross} polarisations of GWs incident from the GRB sky location.
\textsc{X-Pipeline} then constructs linear combinations of the time-frequency maps of the detector data, and a clustering algorithm is used to search for groups of loud pixels in the combined map.
For clusters with significent excess energy, the signal strength in the $N$-dimensional space of detector data streams is projected along directions that are parallel and perpendicular to the projection of a true GW signal from the GRB sky location.
A cluster whose signal energy lies in a direction orthogonal to a true GW signal (for example, with larger amplitude in the least sensitive detector) is rejected by the analysis; these \emph{coherent consistency checks} are described in detail in \cite{Sutton:2009gi, Was:2012xp}.
Clusters (now referred to as events) that survive the consistency checks are ranked by their likelihood to be a circularly polarised GW signal, measured using a Bayesian detection statistic described in \cite{Searle:2009, Was:2012xp}.
Before considering the events found in the on-source region, the off-source results are examined for data quality issues and sensitivity to simulated GW waveforms.

To estimate the rate of background events, the analysis of the off-source data is repeated many hundreds of times, with unphysical \emph{time-slides} of $> 3$\,s applied to the time-series from one of the detectors.
This technique robustly estimates the frequency at which random noise fluctuations in the detectors may appear to be a true GW signal with large amplitude in the on-source window.
A typical search with \textsc{X-Pipeline} will perform $\mathcal{O}(10^{3})$ time-slides on the data in the off-source window, enough to quantify the rate of background events to a \emph{false-alarm rate} (FAR) below $10^{-6}\,\mathrm{Hz}$.
Repeating the analysis for hundreds of time-slides is the most computationally intensive portion of the search.

To determine if a GW is present in the data, the loudest on-source event is compared to the distribution of off-source events.
The \emph{false-alarm probability} (FAP), or \emph{p-value} of this event is defined as the fraction of off-source events with equal or greater significance; this is an empirical measure of the probability of obtaining such an event in the on-source, under the null hypothesis.
Events with $p < 0.01$ are followed up with detailed investigations to determine if the events can be associated with non-GW noise artefacts in the detectors.

\subsection{Sky Location Uncertainty}

The sky localisations of GRBs detected by the \emph{Fermi} GBM can have uncertainty regions covering hundreds of square degrees, depending on the gamma-ray flux and energy spectrum \cite{Briggs:2009dd}.
In a coherent search for GWs associated with GRBs, performing the analysis using an incorrect sky location can reduce the significance of a GW signal in two ways.

First, the analysis will incorrectly estimate the sensitivity of each detector to GWs from the sky location of the GRB.
This can result in loss of coherent signal energy when the time-frequency maps from each detector are combined.
Over most of the sky, the antenna factors for GW observatories change slowly as a function of sky location, usually a few percent over a few degrees.
We have performed empirical tests of the robustness of our coherent detection statistic to variations in sky localisation of several degrees, and for the majority of positions on the sky the loss of signal is of order a few percent.
We conclude that this effect is not large enough to alter the results of our search.

Second, and more significantly, an error in the sky location will lead to an erroneous time-shift of the detector data vectors when synchronising the arrival time of a GW signal across detectors.
For pairs of ground-based detectors the difference in arrival times is $\mathcal{O}(10)$ milliseconds, and an error in the sky location of a few degrees could introduce incorrect synchronisations of a millisecond or more.
This results in the misalignment of a GW signal by several periods for waveforms with frequency content above $1\,\mathrm{kHz}$, and when the data vectors are combined the coherent signal energy will be diminished.
In the worst case, the waveform will be shifted by a half-period between the detectors, and the signal will cancel entirely in the coherent summation.

The standard solution in coherent GW searches is to repeat the analysis over a discrete grid of sky positions covering most of the uncertainty region.
The grid step is chosen so that the timing synchronisation error between any position in the sky localisation error box and the nearest analysis grid point is less than $25\%$ of the period for the highest-frequency GW signals included in the search.
For simplicity, the step size is held constant across the search area; for uncertainty regions with radii of $\mathcal{O}(10)$ degrees the variation in the magnitude of the time-of-arrival correction does not change enough to warrant a variable grid spacing.

Previous searches have used regular grids of concentric circles around the best estimate of the source location, covering at least $95\%$ of the sky location probability distribution.
For the \emph{Fermi} GBM, the $68$\% containment radius is typically $2^{\circ}$--$3^{\circ}$ due to statistical effects, and the localisations have additional systematic errors of several degrees.
As a result, the $95\%$ containment region can cover hundreds of square degrees, and a search for GW signals with frequencies larger than a few hundred Hz would require tiling the search area with many hundreds of search points.
At each grid point the coherent signal combination will have to be re-computed using the new time-of-arrival corrections.
The background estimation for a search grid of this size will typically require $\mathcal{O}(10^{4})$ CPU hours, depending on the size of the GRB uncertainty region, the sky location, and the GW detectors included in the search.
Even on computing clusters with thousands of CPU cores\,\footnote{LIGO Data Grid, \newline \href{https://www.lsc-group.phys.uwm.edu/lscdatagrid/}{https://www.lsc-group.phys.uwm.edu/lscdatagrid/}}, the analysis for a single GBM event can take several hours to several days to complete.

\begin{figure}
  \centering
  \includegraphics[width=0.52\textwidth]{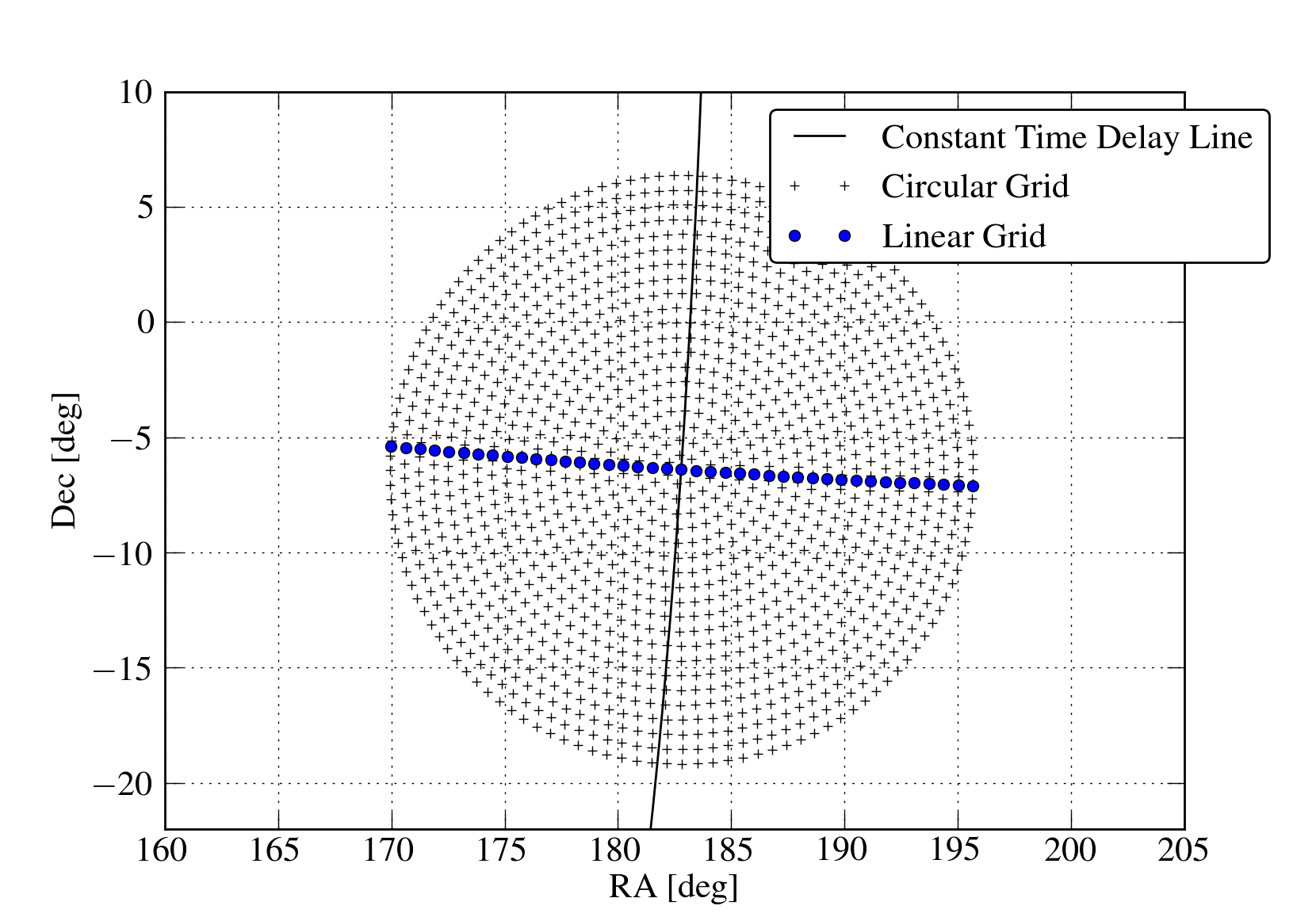}
  \caption{\label{fig:skygrid}Example linear (blue circles) and circular (black crosses) grids for a search for GW signals up to $1792\,\mathrm{Hz}$.
           The localisation for the \emph{Fermi} GBM event GRB\,080906B is shown.
           The linear search grid contains $41$ sky positions, arranged in the direction of the gradient of the time shift been the H2 and G1 detectors.
           The circular grid contains $1324$ sky positions and would require several days to analyse on a massively parallel computing cluster.
           Both search grids cover the $95\%$ containment region for the GBM sky location probability distribution.
           The GBM statistical error for this event is $1.6^{\circ}$.}
\end{figure}

Our solution is to cover the search region with a linear grid, arranged parallel to the maximum gradient of change in the relative time-of-arrival between detectors.
In the case of a 2-detector network, we find that such a pattern is sufficient to capture the dominant source of coherent energy variability as the likelihood is calculated across the GRB uncertainty region.
A comparison of the circular and linear search grids for the \emph{Fermi} event GRB\,080906B is shown in Fig.~\ref{fig:skygrid}.

For two detectors separated by a distance $d$, the difference in time of arrival for a GW is
\begin{equation}
  t = \frac{d \cos{\theta}}{c}
\end{equation}
where $\theta$ is the angle between the inter-detector baseline and the line-of-sight to the GRB, and $c$ is the speed of light.
For a maximum time-delay error tolerance of $dt \leq \alpha$, the corresponding spacing $d\theta$ between grid points is

\begin{equation}
 | d\theta | \leq \frac{2c}{d \sin{\theta}} \alpha
\end{equation}

For our search band of $64\,\mathrm{Hz}$--$1792\,\mathrm{Hz}$\,\footnote{The low frequency limit was chosen to match previous analyses for which the data conditioning has been well tested.
The high-frequency limit is the Nyquist frequency of the detector data ($2048\,\mathrm{Hz}$) minus the widest frequency resolution used in the search ($256\,\mathrm{Hz}$).}, we choose $\alpha=0.14$\,ms, equal to $25\%$ of a cycle at $1792\,\mathrm{Hz}$.
The extent of the linear grid is determined by the $95\%$ containment radius for the given GRB localisation.
For events localised by the \emph{Swift} BAT \cite{Gehrels:2004gu}, we use a search grid of a single point.
For events localised by the \emph{Fermi} GBM \cite{Meegan:2009fe}, we use $1.65\,\sigma_{stat+sys}$, where $\sigma_{stat}$ is the GBM statistical error for the GRB (typically $2^{\circ}$--$3^{\circ}$), and $\sigma_{sys}$ is a $7.5^{\circ}$ systematic error \cite{Connaughton:2011ut, Hurley:2013ip, Singer:2013di}.
The $1.65\,\sigma_{stat+sys}$ uncertainty radius corresponds to $95\%$ containment for a von Mises-Fisher distribution on a sphere \cite{Briggs:2009dd}.

For a handful of GRBs, direct comparisons were made between the linear grid and the full circular tiling, by calculating the sensitivity of the search to simulated GW signals with sky positions distributed across the $95\%$ containment region.
The results for the two methods were nearly identical.
Using the linear grid, the signal amplitudes required for detection were within a few percent of those obtained by the same analysis using the circular grid.
Furthermore, the analysis using the linear grid was completed in a fraction of the time required for the circular grid, and typically required $\mathcal{O}(10^{3}$) or fewer CPU hours, depending on the detectors used in the analysis.
Using computing clusters with thousands of CPU cores, it was possible to analyse some GRBs localised by the GBM in less than three hours.

\subsection{Sky Localisation Using the Linear Grid}

One of the primary goals of GW searches is the prompt localisation of the source sky location, for follow-ups by electromagnetic (EM) astronomers.
Currently, very few GRBs detected by the \emph{Fermi} GBM are examined for optical counterparts, due to the resources necessary to search an uncertainty region of hundreds of square degrees.
The detection of a GW signal associated with a GRB will be of tremendous interest to the astronomical community, and any improvement of the GBM localisation will increase the chances that astronomers could detect an optical, radio, or X-ray counterpart to the gamma-ray and GW signal.

\begin{figure}
  \centering
  \includegraphics[width=0.48\textwidth]{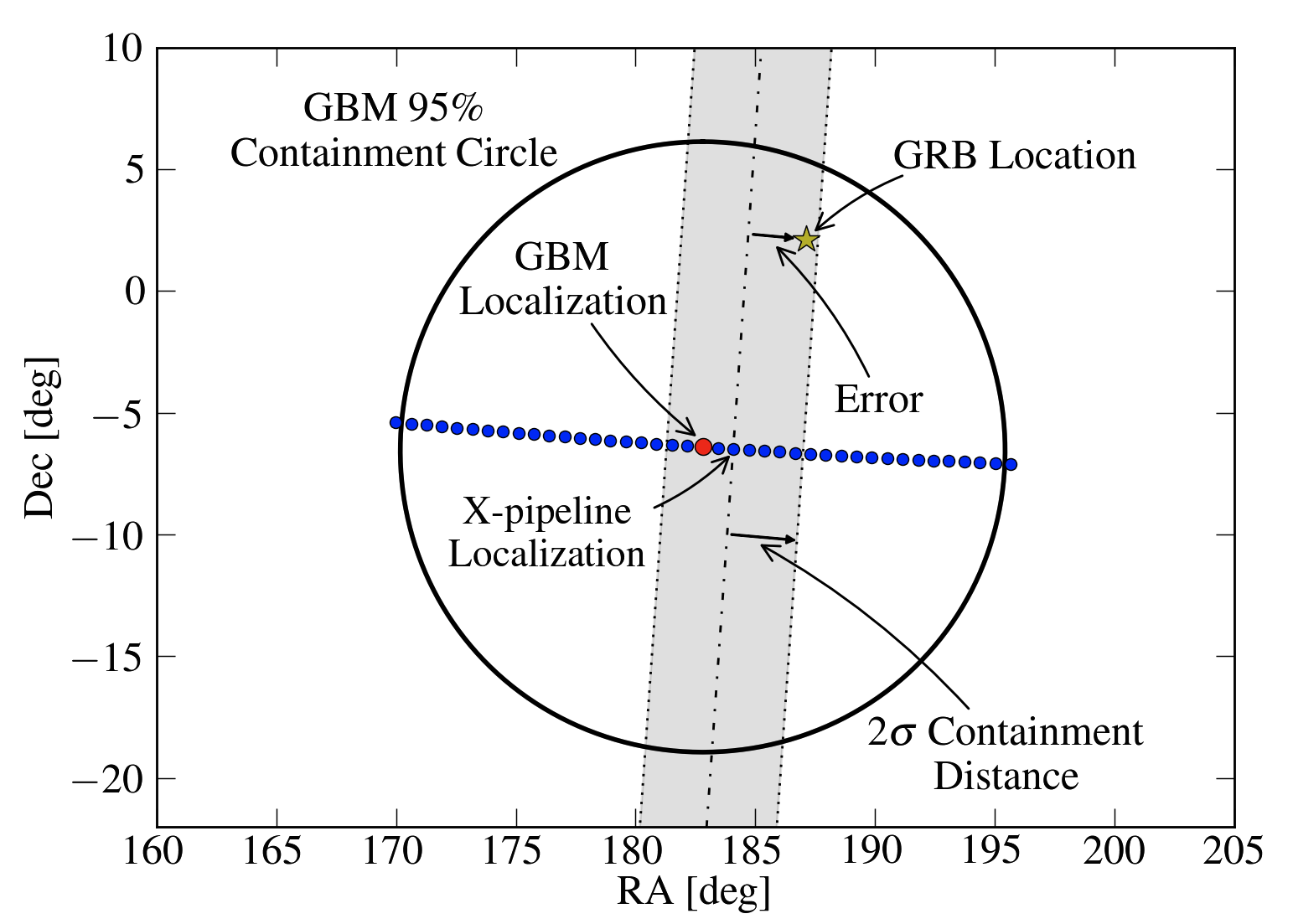}
  \caption{\label{fig:recon_cartoon} Sky localisation using the linear search grid.
           The \textsc{X-Pipeline} analysis will localise a detected GW signal to the point on the linear grid closest to the true injected signal, within errors due to noise fluctuations.
           The containment distance is a function of the frequency and duration of the GW signal, and can be empirically measured for each GRB as part of the GW analysis pipeline; no additional processing time is required.
           In this example, the 2\,$\sigma$ containment distance is about $2.5^{\circ}$.
           The coordinates of the shaded region can be disseminated to EM astronomers for follow-up by wide-field telescopes.}
\end{figure}

If a GW signal is detected in the on-source window, \textsc{X-Pipeline} can localise the source to within a few degrees along the axis of the linear grid, depending on the frequency content of the signal and its duration; this is illustrated in Fig.~\ref{fig:recon_cartoon}.
The sky location of the event is chosen as the point on the search grid that maximises the coherent energy in the detected GW signal.
For a search using data from two widely-separated detectors, the localisation from the GW signal is limited to an annulus on the sky, encircling the line connecting the two detectors \cite{Asai:2013lo}; in principle, this localisation cannot be improved upon in the direction perpendicular to the search grid.
For searches using data from three or more widely-separated detectors, a GW signal can be localised in both dimensions, and in this case the computational cost of the full circular tiling may be worthwhile for event follow-ups.
Efforts to characterise \textsc{X-Pipeline}'s ability to localise signals using three or more detectors are ongoing.

\begin{figure}
  \centering
  \includegraphics[width=0.48\textwidth]{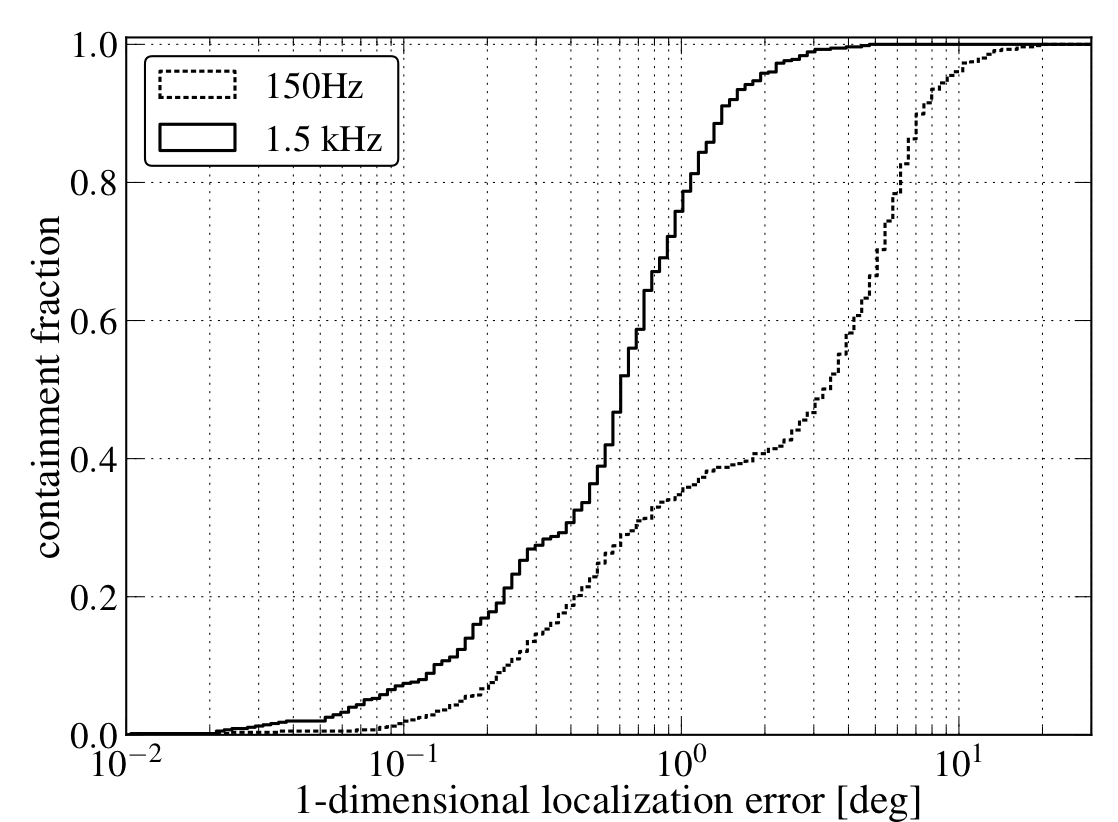}
  \caption{\label{fig:recon_errors} Errors in sky localisation reconstruction, for a simulated GBM event in real GW detector data with the H1-V1 detector network.
           For each frequency, six hundred sine-Gaussian waveforms with quality factor $Q=9$ were analysed using a linear grid covering a $16.5^{\circ}$ uncertainty region with 59 grid points.
           The sky location of each simulation was jittered following a Fisher distribution with $\sigma=10^{\circ}$, and the amplitude was chosen to correspond to the $90\%$ detection threshold.
           The one-dimensional error in the reconstructed location was measured along the axis of the search grid.
           High-frequency waveforms provide greater sensitivity to the time-of-arrival correction across the search grid.}
\end{figure}

As a demonstration of the localisation accuracy, the observed errors for 600 simulated short-duration GW signals with central frequencies of $150\,\mathrm{Hz}$ and $1.5\,\mathrm{kHz}$ are shown in a cumulative histogram in Fig.~\ref{fig:recon_errors}.
The localisation of the high-frequency signals is superior to that of the low-frequency signals due to the greater phase sensitivity to variations in sky position.
For signals at $1.5\,\mathrm{kHz}$, $95\%$ of the simulations were reconstructed to less than $2.5^{\circ}$ along the axis of the linear grid; this provides an empirical measurement of the $2\,\sigma$ uncertainty in \textsc{X-Pipeline}'s localisation.

\section{GRB Sample \& Search Results}
\label{sec:results}

We have applied this technique in a search for GW signals associated with GRBs, using data from the GEO\,600, LIGO, and Virgo observatories.
The GRB events were obtained from the gamma-ray burst coordinates network (GCN) \cite{Barthelmy:2008gc}, supplemented by the \emph{Swift} and \emph{Fermi} on-line catalogues\,\footnote{\href{http://heasarc.gsfc.nasa.gov/W3Browse/fermi/fermigbrst.html}{http://heasarc.gsfc.nasa.gov/W3Browse/fermi/fermigbrst.html}, \newline \href{http://swift.gsfc.nasa.gov/archive/grb\_table}{http://swift.gsfc.nasa.gov/archive/grb\_table} }, as well as the published \emph{Fermi} four-year catalogue \cite{Fermi4yr:2014}.
Most of the GRBs in our sample were detected by \emph{Swift} and \emph{Fermi}; a few of the GRBs were detected by other space borne experiments such as \emph{INTEGRAL} \cite{Winkler:2003in}, \emph{AGILE} \cite{Feroci:2007sa}, or \emph{MAXI} \cite{Matsuoka25102009}.

We analyse GRBs which were observed when GEO\,600 plus one other observatory was taking science-quality data.
For the LIGO and Virgo interferometers, ``science-mode'' is a rigorous definition, and identifies times when the detector configuration is stable and the interferometer is operating in a resonant, low-noise state.
Unbroken intervals of ``science mode'' operation are referred to as ``science segments''; these may last from several minutes to many hours, depending on the environmental conditions and the schedule of instrument upgrades.
Incremental configuration changes are sometimes made in the periods between science segments.
GEO\,600 has no strictly-defined ``science mode'', and collects data on an opportunistic basis between commissioning activities.
In this so-called ``Astrowatch'' operation, efforts are made to collect as much calibrated data as possible given the constraints of commissioning and improvements to the detector.

In our search, no distinction is made between short GRBs and long GRBs, and the analysis is performed without regard to the observed GRB fluence or redshift (if known).
Data segments from GW detectors which are flagged as being of poor quality are excluded from the analysis, and GRBs for which there is insufficient data surrounding the GRB event time are not analysed.
We discard the analysis results of GRBs that are determined to have exceptionally high rates of background events or exceptionally poor sensitivity to GWs from the sky location of the GRB; this can result from, for example, sources of environmental or instrumental noise at the time of the GRB \cite{S5Glitch:2008, LIGODQS6:2010, VirgoDQ:2012, SmithHVeto:2011, TomokiUPV:2010}, or a GRB sky location that includes one of the sensitivity null points of the detectors.
Finally, the sensitivity of the GEO\,600 detector can change by $20\%$ or more at frequencies $>1\,\mathrm{kHz}$, depending on whether or not squeezed light states are being injected.
A change in sensitivity of this magnitude may bias the background estimation if it occurs partway through the off-source window around a GRB.
In this analysis, no GRB off-source (or on-source) windows include times when GEO\,600 changed from a squeezing to non-squeezing state, or vice versa.

In the epoch considered for our search (Feb 4\super{th} 2006 to Nov $3\super{rd}$ 2011), there were $152$ GRBs with sufficient science data to analyse.
For 130 GRBs the results of the background estimation demonstrated good sensitivity to potential GW signals, and for each of these GRBs we calculate the p-value for the loudest event in the on-source.
Three GRBs in our sample had on-source events with $p<0.01$:

\begin{itemize}

\item GRB\,060502A, a \emph{Swift} BAT detection with $\mathrm{T_{90}} = 28.4$\,s and an observed redshift of $z=1.51$ \cite{Cucchiara:2006}, was analysed using data from the L1 and G1 detectors.
There were three significant events in the on-source window.
An examination of the data quality around the time of the GRB revealed non-stationary noise in the L1 detector, associated with increased ground motion due to a magnitude $5.0$ earthquake in Costa Rica.
All three on-source events occurred during a segment of time that was identified as likely to experience an increased rate of transient signals, due to larger than normal seismic vibrations.
Since $35\%$ of the on-source window for this GRB was flagged as having elevated ground motion, we veto the three events and do not include this GRB in the cumulative results.

\item GRB\,080816A, a \emph{Fermi} GBM detection with $\mathrm{T_{90}} = 4.6$\,s, was analysed using data from the H2 and G1 detectors.
There was one significant on-source event, with $p=0.001$.
A signal processing algorithm revealed multiple instrumental channels in the H2 detector with non-stationary noise at the time of the event.
The nature of the instrumental noise is not understood, but the signal in the GW channel is unlikely to be of astrophysical origin.
No redshift observations are available for this GRB.

\item GRB\,090712A, a \emph{Swift} BAT detection with $\mathrm{T_{90}} = 145$\,s, was analysed using data from the G1 and V1 detectors.
There was one significant on-source event, with $p=0.003$.
While we find no plausible instrumental or environmental cause for the event, the observed p-value for this GRB is not significant in a data set containing \numgrbs{} GRBs.
No redshift observations are available for this GRB.

\end{itemize}

\begin{figure}
  \centering
  \includegraphics[width=0.45\textwidth]{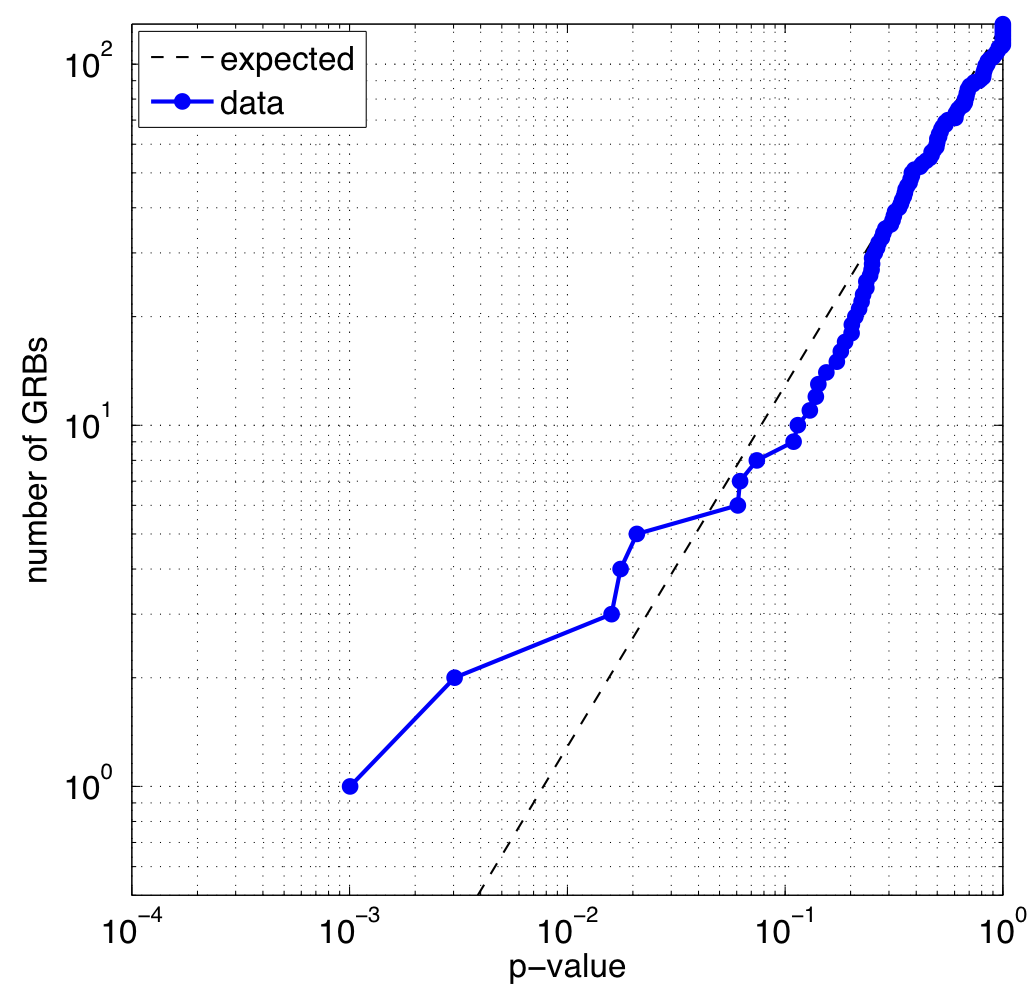}
  \caption{\label{fig:FAP_distribution}Cumulative p-value distribution from the analysis of \numgrbs{} GRBs.
           The dashed line gives the expected distribution under the null hypothesis.
           The most significant on-source event, with $p=0.001$, is associated with GRB\,080816A; a study of detector data at the time of the event yields potential instrumental causes for the signal in the GW channel.
           The probability that our cumulative distribution is due to background is \bkprob, which indicates that the data is consistent with no sub-threshold GW events being present.}
\end{figure}

The distribution of p-values for the most significant event found in the on-source window for each of the \numgrbs{} GRBs is shown in Fig.~\ref{fig:FAP_distribution}.
To test the sample of GRBs for a population of sub-threshold GW signals, we use a weighted binomial test to check that the distribution of p-values is compatible with the uniform distribution expected from the null hypothesis (see Appendix.~A of \cite{Abadie:2012br} for details).
The test yields a background probability of \bkprob{}, which indicates that the distribution is consistent with no GW events being present.

As part of the analysis, we measured the sensitivity of the search to simulated GW signals, as a function of amplitude.
For this search, we simulate GWs from GRBs using circularly polarised sine-Gaussians (SGs) with quality factor $Q=9$.
These \emph{ad hoc} waveforms model the GW emission of a rigidly rotating quadrupolar mass moment with a Gaussian-shaped amplitude evolution in time, and are the standard examples used for estimating the sensitivity of GW analyses to unmodelled, short-duration signals.
We marginalise over systematic errors in sensitivity and phase between detectors by ``jittering'' the simulated waveforms in amplitude and central time before adding them to the detector data; the magnitude of the jitter is Gaussian-distributed with a width proportional to the calibration errors of each detector.
Furthermore, for GRBs detected by the GBM, the sky positions of the simulated waveforms are distributed according to the systematic uncertainties of the GBM detector \cite{Connaughton:2011ut}.
This sky position jittering is performed across the entire uncertainty region, and is not restricted to the axis of the linear search grid.

For each GRB we calculate the total amplitude in GW-induced strain that would result in a detection for $90\%$ of the simulated signals; these $90\%$ upper limits are given in Tab.~\ref{tab:geo_recovered_results}.
In terms of GW strain amplitude, the median $90\%$ upper limit for our GRB sample was $2.8 \times 10^{-21}\,\mathrm{Hz^{-1/2}}$ for circularly-polarised SG signals at $500\,\mathrm{Hz}$ and $3.4 \times 10^{-21}\,\mathrm{Hz^{-1/2}}$ at $1\,\mathrm{kHz}$.
Signals with frequencies below $300\,\mathrm{Hz}$ were typically only detectable at very large amplitude; this is due to the sensitivity of the GEO\,600 detector, and to the choice of coherent cut thresholds, which were tuned to minimise the effect of nonstationary noise in GEO at low frequencies.

For a fixed GW emission energy, we calculate the lower limit on the distance to the GRB using:
\begin{equation}
D_{\mathrm{excl}} = \sqrt{\frac{5}{2}} \, \sqrt{\frac{\mathrm{G}}{\pi^2 \, \mathrm{c}^3}} \, \frac{\sqrt{\mathrm{E_{GW}}}}{f_0 \, \mathrm{h_{rss}}}
\end{equation}
where $E_{GW}$ is the energy released by the GRB central engine in GWs, $G$ is Newton's constant, $f_0$ is the central frequency of the rigid rotator model, and $h_{rss}$ is the strain amplitude upper limit observed by the search.
The factor of $\sqrt{5/2}$ arises from the beamed GW emission from a rotating quadrupolar system, viewed on-axis \cite{SuttonBurst:2013}.

\begin{figure}
  \centering
  \includegraphics[width=0.5\textwidth]{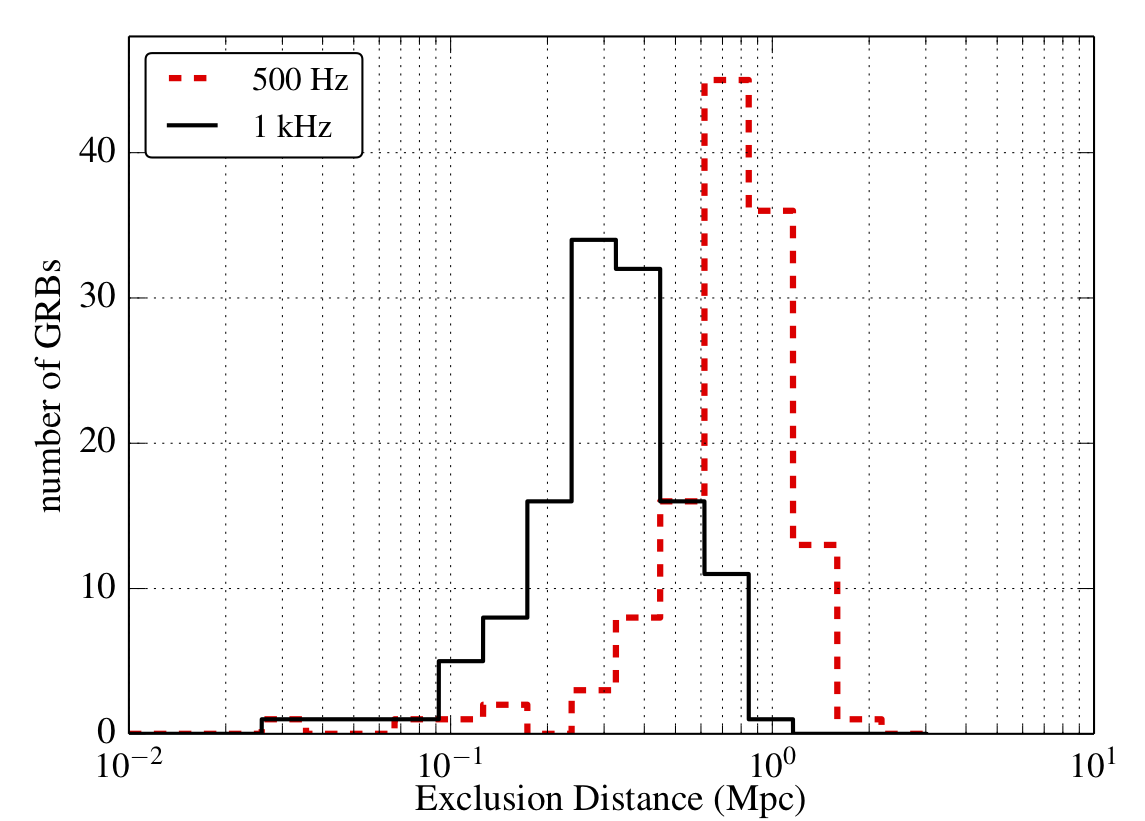}
  \caption{\label{fig:MPC_distribution}Exclusion distances for 500\,Hz and 1\,kHz CSG waveforms for \numgrbs{} GRBs.
           An optimistic total emitted energy of $10^{-2}\,\mathrm{M_{\odot}c^{2}}$ is assumed.}
\end{figure}

The distribution of exclusion distances for waveforms with central frequencies of $500\,\mathrm{Hz}$ and $1\,\mathrm{kHz}$ for the \numgrbs{} GRBs in this search is shown in Sec.~\ref{fig:MPC_distribution}.
The median exclusion distance for the $500\,\mathrm{Hz}$ and $1\,\mathrm{kHz}$ waveforms are \medexdistlowfreq{} and \medexdisthighfreq{} respectively, where we have assumed the GRB central engine releases $E_{GW} = 10^{-2}\,\mathrm{M_{\odot}c^{2}}$ total energy in GWs.
The distance limits scale with the square root of the assumed emission energy, $E_{GW}^{1/2}$, and for a pessimistic assumption of $E_{GW} = 10^{-8}\,\mathrm{M_{\odot}c^{2}}$ the median exclusion distance becomes \medexdistlowfreqkpc{} at $500$\,Hz.
For comparison, the GRB in our sample with the smallest observed redshift is GRB 080905A, with $z=0.1218$ \cite{Rowlinson11102010} or $D \simeq 590$ Mpc \cite{Wright:2006}.

\section{Summary and Future Work}
\label{sec:summary}

We have performed a search for unmodeled short-duration GW signals associated with \numgrbs{} GRBs using data from the GEO\,600, LIGO and Virgo detectors.
The search covered the frequency range $64\,\mathrm{Hz}$--$1792\,\mathrm{Hz}$, and employed a new technique to analyse the large-uncertainty GRB sky localisations from the \emph{Fermi} GBM.
This search is the first to analyse GBM events for GW signals at frequencies above $500\,\mathrm{Hz}$.
We find no evidence for a GW candidate associated with any of the GRBs in this sample, and a statistical analysis shows no sign of a collective signature of sub-threshold GW events.
We have set upper limits on the GW strain at the Earth, assuming a fixed emission energy in GWs from the GRB central engine.
In general, our upper limits only constrain plausible GW emission from GRBs for sources in the Local Group.

The LIGO and Virgo detectors are currently undergoing a major upgrade, implementing new techniques to greatly increase their sensitivity, and are expected to begin operations in 2015.
An improvement on the upper limits presented here of a factor of twenty or more is likely once the advanced detectors reach their design sensitivity.
Various population studies of GRBs with redshift measurements have predicted that the rate of coincident detection in GW observatories and gamma-ray observatories will be $\mathcal{O}(1)$ per year once advanced LIGO and Virgo reach their design sensitivities \cite{PublicRates:2010, Dietz:2011, Leonor:2009, MetzgerCBCEM:2012, HolzBeaming:2013, Siellez:2014}.

Our analysis demonstrates the potential for extending the search frequency band for GWs associated with GBM events above $1\,\mathrm{kHz}$, and the reduced computational cost of this method will be useful for rapid triggered analyses of GBM events in the era of advanced GW detectors.
In the event of a GW detection with signal content above $1\,\mathrm{kHz}$, our search method can provide improved localisation for GRBs with large uncertainties in sky location, which can be passed on to EM astronomers for follow-up using wide-field telescopes.

\section*{Acknowledgements}
We are indebted to the observers of the electromagnetic events and the Gamma-ray burst Coordinates Network for providing us with valuable data.
The authors gratefully acknowledge the support of the United States National Science Foundation for the construction and operation of the LIGO Laboratory, the Science and Technology Facilities Council of the United Kingdom, the Max-Planck-Society, and the State of Niedersachsen/Germany for support of the construction and operation of the GEO\,600 detector, and the Italian Istituto Nazionale di Fisica Nucleare and the French Centre National de la Recherche Scientifique for the construction and operation of the Virgo detector.
The authors also gratefully acknowledge the support of the research by these agencies and by the Australian Research Council, the International Science Linkages program of the Commonwealth of Australia, the Council of Scientific and Industrial Research of India, the Istituto Nazionale di Fisica Nucleare of Italy, the Spanish Ministerio de Econom\'{i}a y Competitividad, the Conselleria d’Economia Hisenda i Innovaci\'{o} of the Govern de les Illes Balears, the Foundation for Fundamental Research on Matter supported by the Netherlands Organisation for Scientific Research, the Polish Ministry of Science and Higher Education, the FOCUS Programme of Foundation for Polish Science, the Royal Society, the Scottish Funding Council, the Scottish Universities Physics Alliance, the National Aeronautics and Space Administration, the National Research Foundation of Korea, Industry Canada and the Province of Ontario through the Ministry of Economic Development and Innovation, the National Science and Engineering Research Council Canada, the Carnegie Trust, the Leverhulme Trust, the David and Lucile Packard Foundation, the Research Corporation, and the Alfred P. Sloan Foundation.
We also gratefully acknowledge the team of graduate students and scientists who maintained the H2 instrument during the Astrowatch epoch, without whom a substantial fraction of our events would not have been analysed.
This paper has been assigned LIGO Document No.\,LIGO-P1300086.

\clearpage
\onecolumngrid
\section*{Table of GRB results}

\newcolumntype{C}{>{\centering\arraybackslash}p{6em}}

\begingroup
\squeezetable
\begin{longtable*}{l|l|r|r|c|C|C|c}
    \hline
    \multicolumn{1}{c|}{GRB} & \multicolumn{1}{c|}{UTC} & & & & \multicolumn{2}{c|}{$90 \%$ ULs ($\times10^{-21}\,\mathrm{Hz^{-1/2}}$)} & \multicolumn{1}{c}{$\gamma$-Ray} \\
    \multicolumn{1}{c|}{name} & \multicolumn{1}{c|}{Time} & \multicolumn{1}{c|}{RA} & \multicolumn{1}{c|}{Dec.} & \multicolumn{1}{c|}{Network} & \multicolumn{1}{c}{$500$\,Hz} & \multicolumn{1}{c|}{$1$\,kHz} & \multicolumn{1}{c}{Detector} \\
    \hline
    \endfirsthead

    \hline
    \multicolumn{8}{c}{\emph{Continued on next page}}
    \endfoot

    \endlastfoot

    \multicolumn{8}{c}{\tablename\ \thetable{} \emph{continued} } \\
    \hline
    \multicolumn{1}{c|}{GRB} & \multicolumn{1}{c|}{UTC} & & & & \multicolumn{2}{c|}{$90 \%$ ULs ($\times10^{-21}\,\mathrm{Hz^{-1/2}}$)} & \multicolumn{1}{c}{$\gamma$-Ray} \\
    \multicolumn{1}{c|}{name} & \multicolumn{1}{c|}{Time} & \multicolumn{1}{c|}{RA} & \multicolumn{1}{c|}{Dec.} & \multicolumn{1}{c|}{Network} & \multicolumn{1}{c}{$500$\,Hz} & \multicolumn{1}{c|}{$1$\,kHz} & \multicolumn{1}{c}{Detector} \\
    \hline
    \endhead

    060424  &  04:16:19 & $ 0\super{h} 29\super{m} 26\super{s}$ &  $ 36^{\circ} 47^{'}$ & G1L1 & $1.46$ & $3.01$ & BAT \\
    060512  &  23:13:20 & $ 13\super{h} 02\super{m} 58\super{s}$ &  $ 41^{\circ} 13^{'}$ & G1L1 & $1.48$ & $2.53$ & BAT \\
    060522  &  02:11:18 & $ 21\super{h} 31\super{m} 49\super{s}$ &  $ 2^{\circ} 53^{'}$ & G1L1 & $1.93$ & $2.35$ & BAT \\
    060602A  &  21:32:12 & $ 9\super{h} 58\super{m} 19\super{s}$ &  $ 0^{\circ} 18^{'}$ & G1L1 & $2.55$ & $4.39$ & BAT \\
    060604  &  18:19:00 & $ 22\super{h} 28\super{m} 54\super{s}$ &  $ -10^{\circ} 56^{'}$ & G1L1 & $1.37$ & $2.58$ & BAT \\
    060708  &  12:15:59 & $ 0\super{h} 31\super{m} 17\super{s}$ &  $ -33^{\circ} 45^{'}$ & G1L1 & $1.61$ & $2.88$ & BAT \\
    060801  &  12:16:15 & $ 14\super{h} 11\super{m} 56\super{s}$ &  $ 16^{\circ} 59^{'}$ & G1L1 & $3.95$ & $3.37$ & BAT \\
    060929  &  19:55:01 & $ 17\super{h} 32\super{m} 35\super{s}$ &  $ 29^{\circ} 50^{'}$ & H2G1 & $2.22$ & $4.93$ & BAT \\
    061110B$^{\dagger}$  &  21:58:45 & $ 21\super{h} 35\super{m} 38\super{s}$ &  $ 6^{\circ} 52^{'}$ & G1L1 & $2.91$ & $7.42$ & BAT \\
    070328  &  03:53:53 & $ 4\super{h} 20\super{m} 27\super{s}$ &  $ -34^{\circ} 04^{'}$ & G1L1 & $1.42$ & $2.21$ & BAT \\
    070406  &  00:50:38 & $ 13\super{h} 15\super{m} 52\super{s}$ &  $ 16^{\circ} 28^{'}$ & G1L1 & $1.37$ & $2.31$ & BAT \\
    070509  &  02:48:27 & $ 15\super{h} 51\super{m} 35\super{s}$ &  $ -78^{\circ} 39^{'}$ & G1L1 & $1.49$ & $2.13$ & BAT \\
    070517  &  11:20:58 & $ 18\super{h} 30\super{m} 14\super{s}$ &  $ -62^{\circ} 18^{'}$ & H2G1 & $2.60$ & $4.12$ & BAT \\
    070925  &  15:52:32 & $ 16\super{h} 52\super{m} 52\super{s}$ &  $ -22^{\circ} 02^{'}$ & G1V1 & $3.01$ & $3.98$ & IBIS \\
    080207$^{\dagger}$  &  21:30:21 & $ 13 \super{h} 50 \super{m} 03 \super{s}$ &  $ 7 ^{\circ} 29 ^{'}$ & G1V1 & $4.09$ & $11.1$ & BAT \\
    080229A  &  17:04:59 & $ 15 \super{h} 12 \super{m} 52 \super{s}$ &  $ -14 ^{\circ} 41 ^{'}$ & H2G1 & $2.21$ & $4.10$ & BAT \\
    080303$^{\dagger}$  &  09:10:35 & $ 7 \super{h} 28 \super{m} 11 \super{s}$ &  $ -70 ^{\circ} 13 ^{'}$ & H2G1 & $3.07$ & $7.78$ & BAT \\
    080319A  &  05:45:42 & $ 13 \super{h} 45 \super{m} 22 \super{s}$ &  $ 44 ^{\circ} 04 ^{'}$ & H2G1 & $2.07$ & $4.02$ & BAT \\
    080319B  &  06:12:49 & $ 14 \super{h} 31 \super{m} 40 \super{s}$ &  $ 36 ^{\circ} 17 ^{'}$ & H2G1 & $2.17$ & $3.13$ & BAT \\
    080319C  &  12:25:56 & $ 17 \super{h} 16 \super{m} 01 \super{s}$ &  $ 55 ^{\circ} 23 ^{'}$ & H2G1 & NaN & $3.09$ & BAT \\
    080328  &  08:03:04 & $ 5 \super{h} 21 \super{m} 58 \super{s}$ &  $ 47 ^{\circ} 31 ^{'}$ & H2G1 & $2.63$ & $4.28$ & BAT \\
    080330  &  03:41:16 & $ 11 \super{h} 17 \super{m} 05 \super{s}$ &  $ 30 ^{\circ} 36 ^{'}$ & G1V1 & $5.72$ & $30.6$ & BAT \\
    080405  &  09:18:55 & $ 10 \super{h} 50 \super{m} 23 \super{s}$ &  $ -4 ^{\circ} 15 ^{'}$ & H2G1 & $1.95$ & $3.07$ & BAT \\
    080411  &  21:15:32 & $ 2 \super{h} 31 \super{m} 50 \super{s}$ &  $ -71 ^{\circ} 17 ^{'}$ & H2G1 & $2.07$ & $3.15$ & BAT \\
    080514B  &  09:55:56 & $ 21 \super{h} 31 \super{m} 16 \super{s}$ &  $ 0 ^{\circ} 44 ^{'}$ & H2G1 & $3.02$ & $15.8$ & AGILE \\
    080515  &  06:01:13 & $ 0 \super{h} 12 \super{m} 36 \super{s}$ &  $ 32 ^{\circ} 34 ^{'}$ & H2G1 & $3.16$ & $5.38$ & BAT \\
    080524  &  04:13:00 & $ 17 \super{h} 54 \super{m} 04 \super{s}$ &  $ 80 ^{\circ} 08 ^{'}$ & H2G1 & $2.03$ & $2.81$ & BAT \\
    080603A  &  11:18:15 & $ 18 \super{h} 37 \super{m} 37 \super{s}$ &  $ 62 ^{\circ} 44 ^{'}$ & H2G1 & $2.38$ & $3.14$ & IBIS \\
    080702A  &  11:50:43 & $ 20 \super{h} 52 \super{m} 14 \super{s}$ &  $ 72 ^{\circ} 16 ^{'}$ & H2G1 & $3.18$ & $4.18$ & BAT \\
    080703  &  19:00:13 & $ 6 \super{h} 47 \super{m} 16 \super{s}$ &  $ -63 ^{\circ} 12 ^{'}$ & H2G1 & $2.53$ & $3.77$ & BAT \\
    080717A  &  13:02:35 & $ 9 \super{h} 49 \super{m} 12 \super{s}$ &  $ -70 ^{\circ} 00 ^{'}$ & H2G1 & $1.78$ & $2.98$ & GBM \\
    080816A  &  12:04:18 & $ 10 \super{h} 24 \super{m} 48 \super{s}$ &  $ 42 ^{\circ} 36 ^{'}$ & H2G1 & $3.01$ & $4.97$ & GBM \\
    080830A  &  08:50:16 & $ 10 \super{h} 40 \super{m} 23 \super{s}$ &  $ 30 ^{\circ} 48 ^{'}$ & H2G1 & $4.34$ & $25.9$ & GBM \\
    080905A  &  11:58:55 & $ 19 \super{h} 10 \super{m} 40 \super{s}$ &  $ -18 ^{\circ} 51 ^{'}$ & H2G1 & $2.26$ & $4.73$ & BAT \\
    080905C  &  13:41:29 & $ 6 \super{h} 27 \super{m} 35 \super{s}$ &  $ -69 ^{\circ} 48 ^{'}$ & H2G1 & $2.99$ & $5.04$ & GBM \\
    080906B  &  05:05:11 & $ 12 \super{h} 11 \super{m} 12 \super{s}$ &  $ -6 ^{\circ} 24 ^{'}$ & H2G1 & $4.20$ & $10.6$ & GBM \\
    080916A  &  09:45:21 & $ 22 \super{h} 25 \super{m} 08 \super{s}$ &  $ -57 ^{\circ} 01 ^{'}$ & H2G1 & $2.73$ & $3.93$ & BAT \\
    081003A  &  13:46:12 & $ 17 \super{h} 29 \super{m} 30 \super{s}$ &  $ 16 ^{\circ} 33 ^{'}$ & H2G1 & $2.81$ & $4.22$ & IBIS \\
    081007  &  05:23:52 & $ 22 \super{h} 39 \super{m} 50 \super{s}$ &  $ -40 ^{\circ} 08 ^{'}$ & H2G1 & $3.76$ & $6.22$ & BAT \\
    081009A  &  03:20:58 & $ 16 \super{h} 41 \super{m} 59 \super{s}$ &  $ 18 ^{\circ} 23 ^{'}$ & H2G1 & $2.58$ & $4.11$ & GBM \\
    081016A  &  06:51:31 & $ 17 \super{h} 02 \super{m} 17 \super{s}$ &  $ -23 ^{\circ} 19 ^{'}$ & H2G1 & $2.77$ & $6.22$ & IBIS \\
    081017B  &  11:22:37 & $ 7 \super{h} 15 \super{m} 59 \super{s}$ &  $ -15 ^{\circ} 12 ^{'}$ & H2G1 & $3.03$ & $4.30$ & GBM \\
    081021A  &  09:33:28 & $ 12 \super{h} 41 \super{m} 12 \super{s}$ &  $ -25 ^{\circ} 36 ^{'}$ & H2G1 & $2.59$ & $3.43$ & GBM \\
    081028B  &  12:55:08 & $ 1 \super{h} 03 \super{m} 59 \super{s}$ &  $ -27 ^{\circ} 12 ^{'}$ & H2G1 & $2.52$ & $3.71$ & GBM \\
    081101  &  11:46:32 & $ 6 \super{h} 23 \super{m} 20 \super{s}$ &  $ -0 ^{\circ} 06 ^{'}$ & H2G1 & $2.02$ & $3.15$ & BAT \\
    081115A  &  21:22:28 & $ 12 \super{h} 42 \super{m} 23 \super{s}$ &  $ 63 ^{\circ} 17 ^{'}$ & H2G1 & $17.0$ & $8.23$ & GBM \\
    081119A  &  04:25:27 & $ 23 \super{h} 6 \super{m} 00 \super{s}$ &  $ 30 ^{\circ} 00 ^{'}$ & H2G1 & $2.92$ & $3.88$ & GBM \\
    081129A  &  03:52:04 & $ 4 \super{h} 12 \super{m} 48 \super{s}$ &  $ -54 ^{\circ} 54 ^{'}$ & H2G1 & $3.23$ & $5.22$ & GBM \\
    081203A  &  13:57:11 & $ 15 \super{h} 32 \super{m} 17 \super{s}$ &  $ 63 ^{\circ} 30 ^{'}$ & H2G1 & $2.05$ & $2.91$ & BAT \\
    081203B  &  13:51:59 & $ 15 \super{h} 15 \super{m} 09 \super{s}$ &  $ 44 ^{\circ} 25 ^{'}$ & H2G1 & $2.26$ & $3.86$ & BAT \\
    081204B  &  12:24:25 & $ 10 \super{h} 03 \super{m} 12 \super{s}$ &  $ 30 ^{\circ} 30 ^{'}$ & H2G1 & $5.81$ & $7.15$ & GBM \\
    081206C  &  23:41:50 & $ 3 \super{h} 37 \super{m} 11 \super{s}$ &  $ -8 ^{\circ} 36 ^{'}$ & H2G1 & $7.47$ & NaN & GBM \\
    081224A  &  21:17:55 & $ 13 \super{h} 26 \super{m} 47 \super{s}$ &  $ 75 ^{\circ} 05 ^{'}$ & H2G1 & $3.53$ & $4.37$ & GBM \\
    081228  &  01:17:40 & $ 2 \super{h} 37 \super{m} 54 \super{s}$ &  $ 30 ^{\circ} 50 ^{'}$ & H2G1 & $2.55$ & $4.23$ & BAT \\
    090123  &  07:51:56 & $ 0 \super{h} 27 \super{m} 10 \super{s}$ &  $ -23 ^{\circ} 30 ^{'}$ & H2G1 & $2.90$ & $6.70$ & BAT \\
    090201  &  17:47:02 & $ 6 \super{h} 08 \super{m} 12 \super{s}$ &  $ -46 ^{\circ} 36 ^{'}$ & H2G1 & $3.50$ & $3.11$ & BAT \\
    090213A  &  05:39:25 & $ 22 \super{h} 02 \super{m} 24 \super{s}$ &  $ -55 ^{\circ} 00 ^{'}$ & H2G1 & $2.03$ & $3.02$ & GBM \\
    090222A  &  04:17:09 & $ 7 \super{h} 54 \super{m} 23 \super{s}$ &  $ 45 ^{\circ} 00 ^{'}$ & H2G1 & $2.59$ & $3.90$ & GBM \\
    090305A  &  05:19:51 & $ 16 \super{h} 07 \super{m} 03 \super{s}$ &  $ -31 ^{\circ} 34 ^{'}$ & H2G1 & $3.44$ & $5.02$ & BAT \\
    090306C  &  05:52:05 & $ 9 \super{h} 07 \super{m} 59 \super{s}$ &  $ 57 ^{\circ} 00 ^{'}$ & H2G1 & $2.63$ & $3.35$ & GBM \\
    090307A  &  03:46:37 & $ 16 \super{h} 19 \super{m} 55 \super{s}$ &  $ -28 ^{\circ} 38 ^{'}$ & H2G1 & $2.12$ & $2.93$ & BAT \\
    090307B  &  03:59:57 & $ 11 \super{h} 30 \super{m} 48 \super{s}$ &  $ -23 ^{\circ} 54 ^{'}$ & H2G1 & $8.70$ & NaN & GBM \\
    090413A  &  02:55:57 & $ 17 \super{h} 45 \super{m} 59 \super{s}$ &  $ -9 ^{\circ} 12 ^{'}$ & H2G1 & $2.43$ & $3.17$ & GBM \\
    090417A  &  13:17:23 & $ 2 \super{h} 19 \super{m} 58 \super{s}$ &  $ -7 ^{\circ} 08 ^{'}$ & H2G1 & $3.08$ & $5.37$ & BAT \\
    090418B  &  09:00:21 & $ 15 \super{h} 03 \super{m} 38 \super{s}$ &  $ 17 ^{\circ} 13 ^{'}$ & H2G1 & $3.01$ & $4.00$ & BAT \\
    090712A & 03:51:05 & $ 4\super{h} 40\super{m} 22\super{s}$ & $ 22^{\circ} 31^{'}$ & G1V1 & $2.21$ & $3.31$ & BAT \\
    090713A & 00:29:28 & $ 18\super{h} 59\super{m} 11\super{s}$ & $ -3^{\circ} 19^{'}$ & G1V1 & $2.85$ & $3.41$ & GBM \\
    090715B & 21:03:15 & $ 16\super{h} 45\super{m} 21\super{s}$ & $ 44^{\circ} 50^{'}$ & G1V1 & $1.59 $ & $1.64$ & BAT \\
    090718B & 18:17:43 & $ 18\super{h} 16\super{m} 24\super{s}$ & $ -36^{\circ} 23^{'}$ & G1H1 & $0.85$ & $1.58$ & GBM \\
    090804A & 22:33:20 & $ 8\super{h} 41\super{m} 36\super{s}$ & $ -11^{\circ} 18^{'}$ & G1V1 & $1.43$ & $1.95$ & GBM \\
    090807B & 19:57:59 & $ 21\super{h} 47\super{m} 35\super{s}$ & $ 7^{\circ} 13^{'}$ & G1V1 & $2.94$ & $5.34$ & GBM \\
    090810A & 18:44:44 & $ 7\super{h} 45\super{m} 43\super{s}$ & $ -17^{\circ} 28^{'}$ & G1V1 & $2.19$ & $3.99$ & GBM \\
    100131A & 17:30:58 & $ 8\super{h} 01\super{m} 36\super{s}$ & $ 16^{\circ} 23^{'}$ & L1G1 & $3.18$ & $7.57$ & GBM \\
    100331A & 00:30:22 & $ 17 \super{h} 24 \super{m} 14 \super{s}$ & $ -58^{\circ} 56^{'}$ & L1G1 & $2.39$ & $2.58$ & IBIS \\
    100417A & 03:59:44 & $ 17\super{h} 25\super{m} 12\super{s}$ & $ 50^{\circ} 23^{'}$ & G1L1 & $26.7$ & $7.05$ & GBM \\
    100510A & 19:27:07 & $ 23 \super{h} 43 \super{m} 12 \super{s}$ & $ -35 ^{\circ} 36 ^{'}$ & L1G1 & $2.91$ & $3.76$ & MAXI \\
    100511A &  00:49:56 & $ 7 \super{h} 17 \super{m} 12 \super{s}$ & $ -4 ^{\circ} 39 ^{'}$ & L1G1 & $5.66$ & $8.70$ & GBM \\
    100528A & 01:48:01 & $ 20 \super{h} 44 \super{m} 24 \super{s}$ & $ 27^{\circ} 48^{'}$ & L1G1 & $2.20$ & $2.43$ & AGILE \\
    100625A & 18:32:28 & $ 1 \super{h} 03 \super{m} 11 \super{s}$ & $ -39^{\circ} 05^{'}$ & L1G1 & $2.56$ & $3.90$ & BAT \\
    100703A & 17:43:37 & $ 0 \super{h} 38 \super{m} 05 \super{s}$ & $ -25^{\circ} 42^{'}$ & L1G1 & $3.13$ & $3.89$ & IBIS \\
    100704A & 03:35:08 & $ 8 \super{h} 54 \super{m} 33 \super{s}$ & $ -24^{\circ} 12^{'}$ & H1G1 & $2.60$ & $3.73$ & BAT \\
    100719C & 19:48:08 & $ 15 \super{h} 25 \super{m} 38 \super{s}$ & $ 18^{\circ} 33^{'}$ & H1G1 & $65.3$ & $17.0$ & GBM \\
    100805A & 04:12:42 & $ 19 \super{h} 59 \super{m} 23 \super{s}$ &  $ 52 ^{\circ} 37 ^{'}$ & L1G1 & $2.23$ & $2.39$ & BAT \\
    100807A & 09:13:13 & $ 3 \super{h} 41 \super{m} 07 \super{s}$ & $ 67^{\circ} 39^{'}$ & L1G1 & $2.04$ & $1.60$ & BAT \\
    100814B & 08:25:26 & $ 8 \super{h} 11 \super{m} 16 \super{s}$ & $ 18^{\circ} 29^{'}$ & L1G1 & $3.63$ & $10.0$ & GBM \\
    100901A & 13:34:10 & $ 1 \super{h} 49 \super{m} 00 \super{s}$ &  $ 22 ^{\circ} 45 ^{'}$ & G1V1 & $7.36$ & $6.16$ & BAT \\
    100906A & 13:49:27 & $ 1 \super{h} 54 \super{m} 47 \super{s}$ & $ 55^{\circ} 38^{'}$ & H1G1 & $1.16$ & $1.16$ & BAT \\
    100907A & 18:01:12 & $ 11 \super{h} 49 \super{m} 09 \super{s}$ & $ -40^{\circ} 37^{'}$ & G1V1 & $6.40$ & $6.15$ & GBM \\
    100915B & 05:49:38 & $ 5 \super{h} 41 \super{m} 34 \super{s}$ &  $ 25 ^{\circ} 05 ^{'}$ & L1G1 & $2.59$ & $3.18$ & IBIS \\
    101008A & 16:43:15 & $ 21 \super{h} 55 \super{m} 31 \super{s}$ & $ 37^{\circ} 03^{'}$ & G1V1 & $4.16$ & $2.79$ & BAT \\
    101017B & 14:51:29 & $ 1 \super{h} 49 \super{m} 52 \super{s}$ & $ -26^{\circ} 33^{'}$ & G1V1 & $4.37$ & $3.37$ & GBM \\
    110604A & 14:49:46 & $ 18 \super{h} 04 \super{m} 00 \super{s}$ & $ 18^{\circ} 28^{'}$ & G1V1 & $4.70$ & $3.18$ & BAT \\
    110605A & 04:23:32 & $ 0 \super{h} 59 \super{m} 47 \super{s}$ & $ 52^{\circ} 27^{'}$ & G1V1 & $1.82$ & $1.49$ & GBM \\
    110610A & 15:21:32 & $ 20 \super{h} 32 \super{m} 49 \super{s}$ & $ 74^{\circ} 49^{'}$ & G1V1 & $2.12$ & $1.93$ & BAT \\
    110616A & 15:33:25 & $ 18 \super{h} 17 \super{m} 48 \super{s}$ & $ -34^{\circ} 01^{'}$ & G1V1 & $2.17$ & $1.60$ & GBM \\
    110618A & 08:47:36 & $ 11 \super{h} 47 \super{m} 13 \super{s}$ & $ -71^{\circ} 41^{'}$ & G1V1 & $2.11$ & $1.61$ & GBM \\
    110624A & 21:44:26 & $ 4 \super{h} 20 \super{m} 04 \super{s}$ & $ -15^{\circ} 57^{'}$ & G1V1 & $2.12$ & $1.79$ & GBM \\
    110625A & 21:08:28 & $ 19 \super{h} 07 \super{m} 00 \super{s}$ & $ 6^{\circ} 45^{'}$ & G1V1 & $2.96$ & $2.15$ & BAT \\
    110626A & 10:44:54 & $ 8 \super{h} 47 \super{m} 38 \super{s}$ & $ 5^{\circ} 33^{'}$ & G1V1 & $3.24$ & $2.32$ & GBM \\
    110629A & 04:09:58 & $ 4 \super{h} 37 \super{m} 28 \super{s}$ & $ 25^{\circ} 00^{'}$ & G1V1 & $3.09$ & $2.86$ & GBM \\
    110702A & 04:29:29 & $ 0 \super{h} 22 \super{m} 28 \super{s}$ & $ -37^{\circ} 39^{'}$ & G1V1 & $5.70$ & $6.70$ & GBM \\
    110706A & 04:51:04 & $ 6 \super{h} 40 \super{m} 19 \super{s}$ & $ 6^{\circ} 08^{'}$ & G1V1 & $5.94$ & $5.64$ & GBM \\
    110709A & 15:24:29 & $ 15 \super{h} 55 \super{m} 34 \super{s}$ & $ 40^{\circ} 55^{'}$ & G1V1 & $2.60$ & $2.07$ & BAT \\
    110709B & 21:32:39 & $ 10 \super{h} 58 \super{m} 40 \super{s}$ & $ -23^{\circ} 28^{'}$ & G1V1 & $3.96$ & $3.18$ & BAT \\
    110709C & 11:06:53 & $ 10 \super{h} 21 \super{m} 31 \super{s}$ & $ 23^{\circ} 07^{'}$ & G1V1 & $2.59$ & $2.02$ & GBM \\
    110709D & 20:40:50 & $ 10 \super{h} 24 \super{m} 50 \super{s}$ & $ -41^{\circ} 47^{'}$ & G1V1 & $3.07$ & $2.87$ & GBM \\
    110710A & 22:53:51 & $ 15 \super{h} 16 \super{m} 21 \super{s}$ & $ 48^{\circ} 23^{'}$ & G1V1 & $1.92$ & $1.61$ & GBM \\
    110716A & 00:25:20 & $ 21 \super{h} 58 \super{m} 43 \super{s}$ & $ -76^{\circ} 58^{'}$ & G1V1 & $2.86$ & $2.48$ & GBM \\
    110722A & 16:39:17 & $ 14 \super{h} 20 \super{m} 14 \super{s}$ & $ 5^{\circ} 00^{'}$ & G1V1 & $2.76$ & $2.03$ & GBM \\
    110729A & 03:25:06 & $ 23 \super{h} 33 \super{m} 33 \super{s}$ &  $ 4 ^{\circ} 58 ^{'}$ & G1V1 & $2.04$ & $1.64$ & GBM \\
    110730B & 15:50:44 & $ 22 \super{h} 20 \super{m} 24 \super{s}$ & $ -2^{\circ} 53^{'}$ & G1V1 & $2.86$ & $2.19$ & GBM \\
    110731A & 11:09:30 & $ 18 \super{h} 42 \super{m} 03 \super{s}$ & $ -28^{\circ} 32^{'}$ & G1V1 & $1.98$ & $1.51$ & BAT \\
    110801A & 19:49:42 & $ 5 \super{h} 57 \super{m} 39 \super{s}$ & $ 80^{\circ} 57^{'}$ & G1V1 & $2.17$ & $1.99$ & BAT \\
    110803A & 18:47:25 & $ 20 \super{h} 01 \super{m} 40 \super{s}$ &  $ -11 ^{\circ} 26 ^{'}$ & G1V1 & $6.02$ & $4.13$ & GBM \\
    110809A & 11:03:34 & $ 11 \super{h} 28 \super{m} 40 \super{s}$ & $ -13^{\circ} 55^{'}$ & G1V1 & $3.91$ & $3.48$ & GBM \\
    110817A & 04:35:12 & $ 22 \super{h} 24 \super{m} 09 \super{s}$ & $ -45^{\circ} 50^{'}$ & G1V1 & $3.47$ & $2.77$ & GBM \\
    110818A & 20:37:49 & $ 21 \super{h} 09 \super{m} 29 \super{s}$ & $ -63^{\circ} 58^{'}$ & G1V1 & $3.55$ & $3.01$ & BAT \\
    110825B & 06:22:11 & $ 16 \super{h} 45 \super{m} 14 \super{s}$ & $ -80^{\circ} 16^{'}$ & G1V1 & $2.35$ & $2.14$ & GBM \\
    110827A & 00:01:52 & $ 10 \super{h} 56 \super{m} 14 \super{s}$ & $ 53^{\circ} 49^{'}$ & G1V1 & $4.11$ & $3.41$ & BAT \\
    110828A & 13:48:15 & $ 7 \super{h} 22 \super{m} 19 \super{s}$ & $ -23^{\circ} 48^{'}$ & G1V1 & $4.45$ & $4.91$ & GBM \\
    110831A & 06:45:27 & $ 23 \super{h} 29 \super{m} 24 \super{s}$ & $ 33^{\circ} 39^{'}$ & G1V1 & $5.34$ & $3.70$ & GBM \\
    110903A & 02:39:55 & $ 13 \super{h} 08 \super{m} 14 \super{s}$ & $ 58^{\circ} 59^{'}$ & G1V1 & $3.02$ & $2.96$ & BAT \\
    110903B & 00:13:06 & $ 10 \super{h} 56 \super{m} 50 \super{s}$ & $ 42^{\circ} 04^{'}$ & G1V1 & $18.8$ & $5.62$ & GBM \\
    110904A & 02:58:16 & $ 23 \super{h} 58 \super{m} 45 \super{s}$ & $ 35^{\circ} 53^{'}$ & G1V1 & $1.86$ & $1.53$ & GBM \\
    110904C & 12:44:19 & $ 21 \super{h} 34 \super{m} 57 \super{s}$ & $ 23^{\circ} 56^{'}$ & G1V1 & $14.3$ & $8.70$ & GBM \\
    111008B & 23:49:01 & $ 14 \super{h} 43 \super{m} 00 \super{s}$ & $ -5^{\circ} 40^{'}$ & G1V1 & $3.35$ & $4.31$ & GBM \\
    111022A & 16:07:04 & $ 18 \super{h} 23 \super{m} 29 \super{s}$ & $ -23^{\circ} 40^{'}$ & G1V1 & $3.88$ & $3.03$ & BAT \\
    111022B & 17:13:04 & $ 7 \super{h} 15 \super{m} 42 \super{s}$ & $ 49^{\circ} 39^{'}$ & G1V1 & $4.29$ & $3.14$ & BAT \\
    111103C & 22:45:06 & $ 13 \super{h} 26 \super{m} 19 \super{s}$ & $ -43^{\circ} 09^{'}$ & G1V1 & $1.58$ & $1.59$ & GBM \\
    \hline
    \caption[Results from GEO recovered gamma-ray burst analyses.]
            {\label{tab:geo_recovered_results}
Information and limits on associated GW emission for each of the analysed GRBs.
The first four columns are: the GRB name in YYMMDD format; the trigger time; and the sky position used for the GW search (right ascension and declination).
The fifth column gives the GW detector network used in the analysis.
Columns six and seven display the upper limits from each GRB: the $90\%$ confidence upper limits on the strain amplitude for circularly polarised $500$\,Hz and $1$\,kHz sine-Gaussian waveforms, in units of $10^{-21} \, \mathrm{Hz^{-1/2}}$.
The last column gives the $\gamma$-ray detector that provided the event time, sky location, sky position uncertainty, and T\sub{90} used for the search (\emph{Swift} BAT, \emph{Fermi} GBM, \emph{INTEGRAL} IBIS, \emph{SuperAGILE}, or \emph{MAXI}).
For three GRBs marked with a $^\dagger$, narrowband non-stationary noise in the GEO\,600 detector at frequencies above $1$\,kHz may have reduced our sensitivity to GW signals.}
\end{longtable*}
\endgroup
\twocolumngrid

\newpage
\bibliographystyle{unsrt}
\bibliography{references}

\end{document}